\documentclass[11pt,a4paper]{article}
\usepackage{jheppub}
\usepackage{amsmath}
\usepackage{amssymb}
\usepackage{graphicx}
\usepackage{subcaption}
\usepackage{slashed}
\usepackage{cases}
\usepackage{tikz-feynman}
\usepackage[toc,page]{appendix}
\usepackage{empheq}

\allowdisplaybreaks

\preprint{OSU-HEP-19-11}

\begin{document}

\title{Neutrino Non-Standard Interactions via Light Scalars in the Earth, Sun, Supernovae and the Early Universe}

\author[a]{K. S. Babu,}
\author[b,a]{Garv Chauhan}
\author[b]{and P. S. Bhupal Dev}

\affiliation[a]{Department of Physics, Oklahoma State University, Stillwater, OK 74078, USA}
\affiliation[b]{Department of Physics and McDonnell Center for the Space Sciences,  Washington University, \\
St. Louis, MO 63130, USA}

\emailAdd{babu@okstate.edu}
\emailAdd{garv.chauhan@wustl.edu}
\emailAdd{bdev@wustl.edu}

\date{\today}

\abstract{Non-standard interactions (NSI) of neutrinos with matter mediated by a scalar field would induce medium-dependent neutrino masses which can modify oscillation probabilities.  Generating observable effects requires an ultra-light scalar mediator. We derive general expressions for the scalar NSI using techniques of quantum field theory at finite density and temperature, including the long-range force effects, and discuss various limiting cases applicable to the neutrino propagation in different media, such as the Earth, Sun, supernovae and early Universe. We also analyze various terrestrial and space-based experimental constraints, as well as astrophysical and cosmological constraints on these NSI parameters, applicable to either Dirac or Majorana neutrinos.  By combining all these constraints, we show that observable scalar NSI effects, although precluded in terrestrial experiments, are still possible in future solar and supernovae neutrino data, and in cosmological observations such as cosmic microwave background and big bang nucleosynthesis data.}

\maketitle

\section{Introduction}
 The discovery of neutrino oscillations implies that at least two of the three neutrinos must have small but non-zero masses~\cite{Tanabashi:2018oca}. The global neutrino oscillation program is now entering a new era, where the known mixing angles and mass-squared differences are being measured with an ever-increasing accuracy. Next-generation of long-baseline oscillation experiments like DUNE are poised to resolve the sub-dominant effects in oscillation data sensitive to the currently unknown oscillation parameters, namely the Dirac $\textit{CP}$ phase, sign of the atmospheric mass-squared difference and the octant of the atmospheric mixing angle. These analyses are usually performed within the $3\times 3$ neutrino mixing scheme under the assumption that neutrinos interact with matter only through the weak interactions mediated by Standard Model (SM) $W$ and $Z$ bosons. On the other hand, the origin of neutrino mass clearly requires some new physics beyond the SM, which often comes with additional non-standard interactions (NSI) of neutrinos with matter fermions (i.e. electrons and/or nucleons). Allowing for these  NSI in neutrino production, propagation and/or detection can in principle change the whole picture and crucially affect the interpretation of the experimental data in terms of the relevant $3\times 3$ oscillation parameters. It is, therefore, of paramount importance to understand all possible kinds of NSI effects, and to see how large these effects could be, while being consistent with other theoretical and experimental constraints. The study of NSI also opens up the possibility of using neutrino oscillations to probe the origin of neutrino mass. 
 
Following the SM interactions of neutrinos with matter via either charged-current (CC) or neutral-current (NC), which after Fierz transformation can be written in the form $(\bar{\nu}_\alpha \gamma^\mu P_L\nu_\alpha)(\bar{f}\gamma_\mu P f)$ (with $f,f'\in \{e,u,d\}$ the matter fermions and $P\in \{P_L,P_R\}$ the chirality projection operators), NSI induced by either a vector or charged-scalar mediator can be parametrized in terms of vector and axial-vector currents~\cite{Wolfenstein:1977ue}: 
\begin{align}
    {\cal L}_{\rm eff}^{\rm V,NC} \ = \ -2\sqrt 2 G_F \sum_{f,P,\alpha,\beta}\varepsilon_{\alpha\beta}^{f,P}(\bar{\nu}_\alpha \gamma^\mu P_L\nu_\beta)(\bar{f}\gamma_\mu P f) \, , \label{eq:VNC} \\
    {\cal L}_{\rm eff}^{\rm V,CC} \ = \ -2\sqrt 2 G_F \sum_{f,P,\alpha,\beta}\varepsilon_{\alpha\beta}^{f,P}(\bar{\nu}_\alpha \gamma^\mu P_L\ell_\beta)(\bar{f}\gamma_\mu P f') \, , \label{eq:VCC}
\end{align}
where $G_F$ is Fermi's constant and the $\varepsilon$ terms quantify the size of the new interactions. The vector components of NSI given by Eq.~\eqref{eq:VNC} and \eqref{eq:VCC} affect neutrino oscillations during 
propagation in matter by providing a new flavor-dependent matter potential. The size of vector NSI is governed by the parameter $\varepsilon\sim g_X^2m_W^2/(g^2m_X^2)$, where $g_X$ and $m_X$ are respectively the coupling and mass of the mediator $X$, and $g$ is the $SU(2)_L$ gauge coupling. There are two possibilities to realize  experimentally observable vector NSI, which require $\varepsilon_{\alpha\beta} \gtrsim 10^{-2}$~\cite{Babu:2019mfe, Coloma:2019mbs}: (i) heavy mediator case with $m_X\sim {\cal O}(100)$ GeV and $g_X\sim {\cal O}(1)$; and (ii) light mediator case with $m_X\ll m_W$ and $g_X\ll 1$ such that $g_X^2/m_X^2 \sim G_F$, while evading the low-energy experimental constraints. For concrete ultraviolet (UV)-complete model realizations, see e.g.~Ref.~\cite{Babu:2019mfe} for the heavy mediator case and Refs.~\cite{Farzan:2015doa, Farzan:2015hkd, Babu:2017olk} for the light mediator case. For a recent review on different aspects of vector NSI, see Ref.~\cite{Dev:2019anc}. For the current global status of the constraints on $\varepsilon$, see Ref.~\cite{Coloma:2019mbs}.

On the other hand, NSI induced by a neutral scalar mediator is no longer composed of vector current as in Eq.~\eqref{eq:VNC} or~\eqref{eq:VCC}, but a scalar interaction  for Dirac neutrinos given by~\cite{Ge:2018uhz}
\begin{align}
    {\cal L}_{\rm eff}^{\rm S} \ = \ \frac{y_f y_{\alpha\beta}}{m_\phi^2}(\bar{\nu}_\alpha \nu_\beta)(\bar{f}f) \, ,
    \label{eq:scalar}
\end{align}
where $y_f$ and $y_{\alpha\beta}$ are respectively the Yukawa couplings of the matter fermion and neutrinos to the scalar mediator $\phi$. This cannot be converted to vector currents, and hence, does not contribute to the matter potential.\footnote{The same is true for tensor NSI of the form $(\bar{\nu}_\alpha\sigma^{\mu\nu}\nu_\beta)(\bar{f}\sigma_{\mu\nu}f)$.} 
Instead, it appears as a medium-dependent correction to the neutrino mass term, with the correction factor $\Delta m_{\nu,\alpha\beta}$ being inversely proportional to the square of the mediator mass. As we will explicitly show below, large enough scalar NSI  effect is possible only for a sufficiently light scalar mediator,\footnote{Eq.~\eqref{eq:scalar} is equally applicable for both light and heavy mediator, since we are dealing with coherent forward scattering of neutrinos with $q^2\to 0$.} since we need $G_{\rm eff}\equiv y_fy_{\alpha\beta}/m_\phi^2\sim 10^{10}G_F$ to have any observable effect for neutrino propagating in Earth with $\Delta m_\nu\sim {\cal O}(0.1 m_\nu)$. Nevertheless, this could potentially lead to significantly different phenomenological consequences in reactor, solar, atmospheric and accelerator neutrino oscillations, as well as for supernovae and early-universe neutrino interactions. 

In this paper, we derive a general formula for evaluating the scalar NSI of the neutrinos which is applicable to different environments. We  perform a systematic study of the scalar NSI in presence of a light scalar mediator $\phi$. We consider both Dirac and Majorana neutrino possibilities. The main objective of our paper is to provide a general field-theoretic derivation of the scalar NSI effect at finite temperature and density, which can be applied to different environments, such as Earth, Sun, supernovae and early Universe. Then we go on to derive various constraints on the couplings in Eq.~\eqref{eq:scalar} as a function of the mediator mass $m_\phi$ from fifth force experiments, solar and supernova neutrino data, stellar cooling constraints from red giants (RG) and horizontal branch (HB) stars, and big bang nucleosynthesis (BBN). We have considered scalar interactions with electrons and nucleons separately to show the differences in the constraints. We find that the fifth force experiments constrain masses of $\phi$ below $0.1$ eV and couplings up to $10^{-24}$. RG/HB stars constrain couplings up to $10^{-12}$ for nucleons and up to $10^{-16}$ for electrons coupling to $\phi$. Bounds from BBN constrain couplings up to $10^{-9}$ for the light scalar mediators. After taking into account all these constraints, we conclude that any prospects of observing scalar NSI in Earth matter have been ruled out, while these effects are still  measurable with future solar neutrino data, supernova neutrino bursts or in cosmological observations of extra relativistic degrees of freedom. 

The rest of the paper is organized as follows: In Sec.~\ref{sec:2}, we present a general field-theoretic derivation of scalar NSI and discuss various limiting cases that are applicable to Earth, Sun, supernovae and early Universe. In Sec.~\ref{sec:LongForce}, we discuss the long-range force effects of a light scalar. In Sec.~\ref{sec:4}, we summarize the current experimental constraints on the Yukawa couplings relevant for scalar NSI as a function of the mediator mass. In Sec.~\ref{sec:TM}, we discuss the thermal mass of the mediator. In Sec.~\ref{sec:QM}, we derive a quantum-mechanical bound on the effective in-medium mediator mass. In Sec.~\ref{sec:discussion}, we present our main results and discussions. In Sec.~\ref{sec:8}, we present a UV-complete  model for scalar NSI. Our conclusions are given in Sec.~\ref{sec:9}. In Appendix~\ref{appendix:1}, we give the detailed derivation of various limiting cases for the scalar NSI discussed in Sec.~\ref{sec:2}. In Appendix~\ref{appendix:3}, we provide details of the calculation of the neutrino self-energy in neutrino background. In Appendix~\ref{appendix:2}, we present the calculation for thermal mass of the scalar mediator. 
\section{Field theoretic origin of scalar NSI}\label{sec:2}

In this section, we derive expressions for medium-dependent neutrino mass and energy when the neutrinos have scalar NSI with matter fermions in the propagating medium. The results derived here are equally applicable for Dirac and Majorana neutrinos.  As we will see later, for observable scalar NSI it will be required that the scalar field is very light, which we assume here.  
Consider the interaction of fermions $f$ and Dirac neutrinos $\nu$ with a light scalar $\phi$, with the relevant interaction terms given by the Lagrangian:
\begin{equation}
    \mathcal{L} \ \supset \ - y_{\alpha\beta} \overline{\nu}_\alpha \phi\nu_\beta - y_f \bar{f}\phi f -  m_{\alpha\beta} \overline{\nu}_\alpha \nu_\beta - \frac{m_\phi^2}{2}\phi^2 \, .
    \label{Dirac}
\end{equation}
In the case of Majorana neutrinos, the relevant Lagrangian has the form:
\begin{equation}
    \mathcal{L} \ \supset \ - \frac{y_{\alpha\beta}}{2} \overline{\nu^c_\alpha} \phi\nu_\beta - y_f \bar{f}\phi f -  \frac{m_{\alpha\beta}}{2} \overline{\nu^c_\alpha} \nu_\beta - \frac{m_\phi^2}{2}\phi^2~.
    \label{Majorana}
\end{equation}
We shall focus primarily on the Dirac neutrinos, but essentially all of our results will apply for Majorana neutrinos as well, provided that the normalization of couplings is as in Eq.~(\ref{Majorana}).  We shall comment on differences when they arise between the two cases.

A neutrino with four-momentum $p^\mu$ propagating through matter obeys the Dirac equation given by:
\begin{equation}
    \left[\slashed{p}-\Sigma(p)\right]\psi \ = \ 0~.
\end{equation}
In a general medium, the self energy $\Sigma$ of the neutrino  gets modified. We apply real time formalism of field theory at finite temperature and density in our derivations, which is manifestly Lorentz covariant~\cite{Das:1997gg}. With  pure scalar interactions of the type given in  Eqs.~(\ref{Dirac}) and (\ref{Majorana}), the neutrino self-energy  takes the general form
\begin{equation}\label{eq:genSelfE}
  \Sigma(p) \ = \ m - (\hat{a}\slashed{p}+\hat{b}\slashed{u}+\hat{d}[\slashed{p},\slashed{u}]),
\end{equation}
where $m$ is the neutrino mass inside the medium, $u^\mu$ is the four-velocity of the medium and $\hat{a},\,\hat{b},\,\hat{d}$ are functions of only two Lorentz scalars, viz., $p^2 $ and $p.u$. In a Lorentz covariant description of field theory at finite temperature and density, one introduces a medium four-velocity vector $u^\mu$ as in Eq.~\eqref{eq:genSelfE} obeying $u^2=1$. In real time formalism of thermal field theory, the finite temperature and density correction to self-energy of a fermion can be calculated with the help of finite temperature Green's function for a free Dirac field \cite{Das:1997gg} (for applications to neutrino propagation in matter see Refs.~\cite{Notzold:1987ik,Pal:1989xs, Langacker:1992xk}):
\begin{equation}
    S_f(p) \ = \ (\slashed{p}+m_f)\left[ \frac{1}{p^2-m_f^2+i\epsilon}+ i \Gamma_f(p)\right] \\
\end{equation}
where
\begin{equation}
    \Gamma_f(p) \ = \ 2\pi \delta(p^2-m_f^2)[n_f(p)\Theta(p_0)+n_{\bar{f}}(p)\Theta(-p_0)] ~.
    \label{eq:2.6}
\end{equation}
Here $\Theta$ is the Heaviside step function and $n_f$ ($n_{\bar{f}}$) is
the Fermi-Dirac distribution function for the fermion (anti-fermion) occupation number of the medium given by
\begin{equation}
     n_{f}(p) \ = \ \frac{1}{e^{(|p.u|- \mu)/T}+1}\, , \qquad 
     n_{\bar{f}}(p) \ = \ \frac{1}{e^{(|p.u|+ \mu)/T}+1}~,
\end{equation}
where $\mu$ is the chemical potential and $T$ is the temperature. Integrating the occupation number  over all possible momentum states yields the total number density of the fermions (or anti-fermions) in the medium:
\begin{equation}
    N_{f(\bar{f})} \ = \ g_f \int \frac{d^3p}{(2\pi)^3}n_{f(\bar{f})}(p) \, .
\end{equation}
Here $g_f$ denotes the number of internal degrees of freedom and is equal to two for electrons, nucleons and neutrinos for the two different spin states. 
\begin{figure}
\centering
\begin{subfigure}[b]{.48\textwidth}
  \centering
  \includegraphics[scale=1.3]{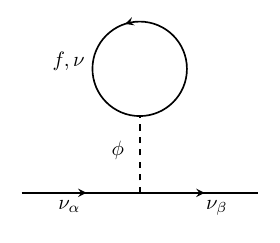}
  \caption{}
  \label{fig:D1}
\end{subfigure}%
\begin{subfigure}[b]{.48\textwidth}
  \centering
   \includegraphics[scale=1.3]{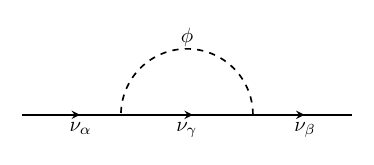}
  \caption{}
  \label{fig:D2}
\end{subfigure}
\caption{Neutrino self-energy diagrams: (a) Tadpole with background of $f$ and $\nu$, and (b) Self-energy in a neutrino background.}
\label{fig:FD}
\end{figure}

\subsection{Neutrino self-energy from tadpole diagram}
The one-loop thermal self energy corrections for the neutrinos arising from Eq.~(\ref{Dirac}) or Eq.~(\ref{Majorana}) are shown in Fig.~\ref{fig:FD}.  We first compute the one-loop neutrino thermal mass correction induced by the tadpole diagram in Fig.~\ref{fig:D1}. The Lorentz-invariant form of $\Sigma$ as given in Eq.~\eqref{eq:genSelfE} can be conveniently evaluated by going to the rest frame of the medium, where the amplitude takes a simple form:
\begin{align}
    -i\Sigma_{\alpha\beta} & \ = \ i y_{\alpha\beta}\:\frac{i}{q^2-m_\phi^2}\int \frac{d^4k}{(2\pi)^4} {\rm Tr}\left[iy_f\:i S_f(k)\right] \, . \label{eq:res1}
\end{align}
In Eq.~\eqref{eq:res1}, we can set $q^2=0$ for the momentum transfer because we are only interested in the coherent forward scattering of neutrinos in matter for the NSI effect. Only retaining the finite temperature and density part of the self-energy, we obtain
\begin{align}
    \Sigma_{\alpha\beta}  & \ = \  \frac{y_{\alpha\beta}y_fm_f}{\pi^2m_\phi^2}\int_0^\infty dk_0\int_0^\infty dk^2\:k\:\delta(k^2-k_0^2+m_f^2) \left[n_f(k_0)+n_{\bar f}(-k_0)\right] \, .   
    \label{int}
\end{align}
Integrating over $k^2$ using the delta function yields the final result:
\begin{align}
    \Sigma_{\alpha\beta}  \ = \  \frac{y_{\alpha\beta}y_fm_f}{\pi^2m_\phi^2}\int_{m_f}^\infty dk_0\:\sqrt{k_0^2-m_f^2} \left[n_f(k_0)+n_{\bar f}(k_0)\right] \ \equiv \ \Delta m_{\nu, {\alpha\beta}}~.
    \label{eq:sNSI}
\end{align}
While deriving Eq.~\eqref{eq:sNSI}, it has been assumed that the background medium contains both fermions and anti-fermions. Thus, Eq.~\eqref{eq:sNSI} is the complete expression for scalar NSI of neutrinos at any finite temperature and density in a background without neutrinos. We have provided details of evaluating the integral of Eq.~(\ref{eq:sNSI}) in various useful limits in Appendix~\ref{appendix:1}.

Note that the scalar NSI of Eq.~(\ref{eq:sNSI}) appears as a medium-dependent mass of the neutrino.  The relevant integral can be evaluated analytically in the high temperature as well as low temperature regimes.  We find:
\begin{numcases}{\Delta m_{\nu, {\alpha\beta}} \ = \ }
\frac{y_f y_{\alpha\beta}}{m_\phi^2}\left(N_f+N_{\bar f}\right) & \textbf{     ($\mu, T\ll m_f$)} \label{eq:c1} \\
 \frac{y_f y_{\alpha\beta}}{m_\phi^2}\frac{m_f}{2}\left(\frac{3}{\pi}\right)^\frac{2}{3} \left(N_f^{2/3}+N_{\bar f}^{2/3}\right) & \textbf{($\mu>m_f\gg T$)} \label{eq:c2} \\
 \frac{y_f y_{\alpha\beta} m_f}{3 \:m_\phi^2}\left(\frac{\pi^2}{12 \:\zeta(3)}\right)^\frac{2}{3} \left(N_f^{2/3}+N_{\bar f}^{2/3}\right) & \textbf{ ($\mu<m_f\ll T$)}~. \label{eq:c3}
\end{numcases}
If the medium does not contain either fermions or anti-fermions of a certain type, the corresponding number density should be set to zero in the final result. If the background has more than one type of fermion, the various contributions should be  added. Eq.~\eqref{eq:c1} for $\mu,T \ll m_f$ is the non-relativistic limit for the scalar NSI expression and matches the result stated in Ref.~\cite{Ge:2018uhz}. It is most useful in case of the Earth and Sun. The limiting case Eq.~\eqref{eq:c2} is useful for relativistic medium backgrounds as with electrons in supernovae. For effect of scalar NSI in the early Universe, Eq.~\eqref{eq:c3} is the most relevant. Detailed application of these results is carried out in Sec.~\ref{sec:discussion}. 

\subsection{Neutrino self-energy in a neutrino background}
\label{sec:2.2}

There is another important diagram that might contribute to the effect of neutrino propagation in a medium, as shown in Fig.~\ref{fig:D2}. This diagram contributes to neutrino self-energy only in media with a neutrino or an anti-neutrino background. This situation is realized in supernovae and early Universe cosmology. Here we derive the contribution of Fig.~\ref{fig:D2} in such backgrounds. 
Again using the real-time formalism of thermal field theory, we can write this contribution for a Dirac neutrino as:
\begin{equation}
    \Sigma_{\alpha\beta}^\nu \ = \ -y_{\beta\gamma} y_{\gamma\alpha} \int \frac{d^4k}{(2\pi)^4} \left(\slashed{k}+\frac{\slashed{p}}{2}+m_\nu\right)\left[ \frac{\Gamma_\phi\left(k-\frac{p}{2}\right)}{\left(k+\frac{p}{2}\right)^2-m_\nu^2} + \frac{\Gamma_\nu\left(k+\frac{p}{2}\right)}{\left(k-\frac{p}{2}\right)^2-m_\phi^2}\right] \, ,
    \label{self-energy}\end{equation}
where $\Gamma_\nu$ is defined in Eq.~\eqref{eq:2.6} and for $\Gamma_\phi$, we have used the finite temperature Green's function for a free bosonic field given by:
\begin{equation}
    S_b(p) \ = \ \left[ \frac{1}{p^2-m_b^2+i\epsilon}- i \Gamma_b(p)\right] \\
\end{equation}
where
\begin{equation}
    \Gamma_b(p) \ = \ 2\pi \delta(p^2-m_b^2)n_b(p)\Theta(p_0) \, ,
\end{equation}
with the Bose-Einstein distribution function given by
\begin{equation}
     n_{b}(p) \ = \ \frac{1}{e^{(|p.u|)/T}-1}~,
\end{equation}
noting that the chemical potential of the real scalar field $\phi$ is zero. 
We have carried out the evaluation of the self energy integral of Eq.~(\ref{self-energy}) in
Appendix \ref{appendix:3}; here we summarize our main results. The contribution of Eq.~(\ref{self-energy}) can be written as 
\begin{equation}
    \Sigma_{\alpha\beta}^\nu \ = \  -\frac{y_{\beta\gamma} y_{\gamma\alpha}}{8\pi^2|\textbf{p}|} J \, ,
    \label{eq:2.19}
\end{equation}
with $J$ identified as the integral of Eq.~(\ref{self-energy}), except for an overall factor, and can be decomposed as
\begin{equation}
J \ = \ a\slashed{p}+b\slashed{u}+c+d[\slashed{p},\slashed{u}]~.
\label{eq:J}
\end{equation}
By taking traces of the integral in Eq.~(\ref{self-energy}) multiplied by $(1,\,\slashed{p},\,\slashed{u},\,\slashed{p}\slashed{u})$, we can solve for the Lorentz scalars  $(a,\,b,\,c,\,d)$. Defining
\begin{equation}
    J_p \ = \ \text{Tr}(J\slashed{p})\, ,\quad J_u \ = \ \text{Tr}(J\slashed{u}) \, , \quad \text{and} \quad J_m \ = \  \text{Tr}(J) \, ,
    \label{eq:Jpum}
\end{equation}
we find
\begin{equation}
    a \ = \  \frac{J_u\:(p.u)-J_p}{4[(p.u)^2-p^2]}
    \, , 
    \quad b \ = \  \frac{J_p\:(p.u)-J_u p^2}{4[(p.u)^2-p^2]} \, ,
    \quad c \ = \ \frac{J_m}{4}\, , \quad \text{  and} \quad d \ = \ 0~.
    \label{eq:abcd}
\end{equation}
It is clear that the coefficient $c$ contributes to the neutrino mass in the medium [cf.~Eq.~\eqref{eq:genSelfE}]. But this effect is negligible in our case, because there is no $1/m_\phi^2$ enhancement. 

There is also a matter potential that is caused by the neutrino self-interactions. 
To arrive at it we examine the pole in the neutrino propagator:
\begin{eqnarray}
i S_\nu^{-1}(p)  \ = \  i (\slashed{p}-\Sigma^\nu)  \ = \  i[\slashed{p}(1-A)- B \slashed{u}] \, ,
\end{eqnarray}
where $A$ and $B$ are matrices in flavor-space, with elements given by 
\begin{equation}
A_{\alpha\beta} \ = \  -\frac{y_{\beta\gamma}y_{\gamma\alpha}}{8 \pi^2 |{\bf p}|} a \, , \qquad 
B_{\alpha\beta} \ = \ -\frac{y_{\beta\gamma}y_{\gamma\alpha}}{8 \pi^2 |{\bf p}|} b~.
\label{eq:AB}
\end{equation}
Since $A$ and $B$ commute, $S_\nu$ can be obtained in terms of $A$ and $B$ as
\begin{eqnarray}
i S_\nu(p) \ = \ i \frac{[(1-A)\slashed{p} - B \slashed{u}]}{\{(1-A) p-B u\}^2}~.
\label{pole}
\end{eqnarray}
We define energy and momentum of the neutrino (in the massless limit) in the rest frame of the medium  as~\cite{Pal:1989xs}
\begin{equation}
{\cal E} \ = \ p.u, \qquad {\cal P} \ = \ \sqrt{{\cal E}^2-p^2}~.
\end{equation}
The pole in the neutrino propagator of Eq.~(\ref{pole}) occurs at energy values given by
\begin{equation}
   {\cal E}_i \ = \ \frac{B_i}{1-A_i} \pm {\cal P}~.
   \label{shift}
\end{equation}
This leads to the modified dispersion relation 
${\cal E}=U{\cal E}_i U^\dag$ (where $U$ is the unitary matrix that diagonalizes $A$ and $B$). 
The energy shift for neutrinos is thus $B/(1-A)$, while the shift in antineutrino energy is $-B/(1-A)$, which are both non-diagonal in the flavor basis [cf.~Eq.~\eqref{eq:AB}].   

For significant regions of the Yukawa couplings $y_{\alpha\beta}$ and $y_f$, the scalar $\phi$ does not get thermalized. In this case, there is no $\phi$ background and the term proportional to $\Gamma_\phi(k-p/2)$ should be set to zero. We present our results here in this case first. The contribution from $\Sigma^\nu_{\alpha\beta}$ can then be written as:
\begin{equation}\label{eq:calcSE}
    \Sigma_{\alpha\beta}^\nu \ = \ -y_{\alpha\gamma} y_{\gamma\beta} \int_{\frac{-p_0}{2}}^\infty dk_0\int \frac{d^3k}{(2\pi)^3}  \frac{(\slashed{k}+\frac{\slashed{p}}{2}+m_\nu)}{\left(k-\frac{p}{2}\right)^2-m_\phi^2}\delta\left[{\left(k+\frac{p}{2}\right)^2-m_\nu^2}\right]n_\nu \left(k_0+\frac{p_0}{2}\right)~.
\end{equation}
We defer the details of evaluating this integral to Appendix~\ref{appendix:3}. Here we present the results in the high temperature limit, assuming that the chemical potential is vanishing. This condition is generally true in the early Universe when neutrinos propagate in a background of neutrinos. Furthermore, we set the neutrino mass to be zero, which is a consistent approximation as the medium-induced $m_\nu$ is proportional to the original $m_\nu$. In the absence of neutrino mass, we can set $p_0^2-|{\bf p}|^2 = 0$. Under these conditions, our results are as follows (see Appendix \ref{appendix:3} for details):
\begin{eqnarray}
a & \ = \ & -\frac{\pi^2 T^2}{24 |{\bf p}|}\left[2-12 \zeta'(-1) - {\rm ln} \left(\frac{16 \pi |{\bf p}|T}{m_\phi^2} \right)  \right] - \frac{T}{4}\, {\rm ln}2 \,\, {\rm ln}\left(\frac{2 \sqrt{2} |{\bf p}| T}{m_\phi^2} \right) \, ,  \nonumber \\
b & \ = \ & \frac{\pi^2 T^2}{12}~.
\end{eqnarray}
Here $\zeta'(-1) = -0.165421$ is the derivative of Riemann zeta function evaluated at argument equal to $-1$.  Using these results along with Eq.~(\ref{shift}), we arrive at the energy shift experienced by the neutrino in a background of neutrinos:
\begin{eqnarray}
\Delta{\cal E}_{+,\alpha\beta} & \ = \ &
-\frac{T^2}{96 |{\bf p}|}\left[
y y^\dagger \left( 1- y y^\dagger \frac{T^2}{192 |{\bf p}|^2}
\left\{2-12 \zeta'(-1) - {\rm ln} \left(\frac{16 \pi |{\bf p}|T}{m_\phi^2}\right)  \right\} \right. \right. \nonumber \\ 
& & \left. \left. - y y^\dagger\frac{T}{32\pi^2|{\bf p}|}\, {\rm ln}2 \,\, {\rm ln}\left(\frac{2 \sqrt{2} |{\bf p}| T}{m_\phi^2} \right) \right)^{-1}\right]_{\alpha\beta}~.
\label{energy-shift}
\end{eqnarray}
Here we have made use of the fact that $yy^\dagger = U D U^\dagger$, where $D$  is a diagonal matrix and $U$ is unitary, obtained the poles in the neutrino propagators in the diagonal basis, and reinserted the unitary matrix in writing Eq.~(\ref{energy-shift}). 
\color{black}
While we do not use these results explicitly in our analyses, these are part of the neutrino scalar NSI which may find use in early Universe cosmology where there is a thermal background of neutrinos.

If the scalar field $\phi$ is also in thermal equilibrium, a similar analysis goes through albeit with some replacements, as can be seen from Eq.~(\ref{self-energy}): $\Gamma_\nu \rightarrow \Gamma_\phi$, $p \rightarrow -p$, with a change in sign of $\slashed{p}$ in the numerator and change of $m_\phi\rightarrow m_\nu$ only in the denominator. These thermal $\phi$ contributions will add to the neutrino self-energy contribution to  $J$ given in Eq.~(\ref{eq:J}).  In particular, the coefficients $J_p,\,J_u,\,J_m$ of Eq.~(\ref{eq:Jpum}) will become $J_p + J_p^\phi,\,  J_u + J_u^\phi,\, J_m + J_m^\phi$, where the new contributions are given in Appendix~\ref{appendix:3}.

\section{Long-range force effects}\label{sec:LongForce}

A light scalar coupling to fermions can lead to  long-range forces. This applies to charged fermions as well as neutrinos propagating through a medium. Even when the neutrino propagates outside of the medium, such long-range forces can affect its propagation. Thus, calculating the neutrino energy using point interactions with a very light mediator does not provide a complete picture.  In this section, we sketch a heuristic derivation to account for these long-range force effects. Long range effects in non-relativistic media have been studied in Refs.~\cite{Wise:2018rnb, Smirnov_2019}. Here, we have extended the analysis for all background media, i.e. both non-relativistic and relativistic cases. This will be especially useful in relativistic media such as in supernovae and in early Universe.  

We use the Euler-Lagrange equations for the Lagrangian in Eq.~\eqref{Dirac} to obtain equations of motion for $\nu$ and $\phi$:
\begin{align}
    (i\slashed{\partial}- m_{\alpha\beta} - y_{\alpha\beta} \phi ) \nu_\beta  & \ = \ 0 \label{eq:NuEOM}\\
    (\partial^2+m_\phi^2)\:\phi -  y_{\alpha\beta} \overline{\nu}_\alpha \nu_\beta- y_f \bar{f} f & \ = \ 0~. \label{eq:PhiEOM}
\end{align}
As can be seen from Eq.~\eqref{eq:NuEOM}, the interaction vertex $y_{\alpha\beta} \overline{\nu}_\alpha \phi\nu_\beta$ leads to an extra contribution to neutrino mass: 
\begin{equation}\label{eq:massC}
    \Delta m_{\nu, {\alpha\beta}} \ = \  y_{\alpha\beta} \: \langle\phi\rangle_{\text{medium}}~.
\end{equation}
To calculate the mass correction for a neutrino propagating in a medium, we will need to calculate the expectation value of the operators at finite temperature and density, appearing in Eqs.~\eqref{eq:NuEOM} and \eqref{eq:PhiEOM}.

\par For a medium in thermal equilibrium with fermion number density $N_f$ and anti-fermion number density $N_{\bar f}$ can be represented as a Fock state $|\Psi\rangle$. This state contains information about particle and anti-particle distribution in different momentum states. Since the system is assumed to be in thermal equilibrium, the fermion and anti-fermion density in each momentum state does not change in time. Thus, we can set $t=0$ and the state $|\Psi\rangle$ is normalized, i.e, $\langle \Psi|\Psi\rangle = 1$. The field operators for the fermion and anti-fermion fields can be written as~\cite{Peskin:1995ev}:
\begin{align}
    f(x) & \ = \ \int \frac{d^3p}{(2\pi)^3} \frac{1}{\sqrt{2E_\textbf{p}}}\; \; \sum_s \left[a^s_\textbf{p}\: u^s(p) \: e^{-ip.x} + b^{s\dagger}_\textbf{p}\: v^s(p) \: e^{ip.x}
    \right] \, , \\
    \bar{f}(x) &\ = \ \int \frac{d^3p}{(2\pi)^3} \frac{1}{\sqrt{2E_\textbf{p}}}\; \; \sum_s \left[b^s_\textbf{p}\:  \:\overline{v}^s(p) e^{-ip.x} + a^{s\dagger}_\textbf{p}\: \overline{u}^s(p) \: e^{ip.x}\right]~.
\end{align}
We need to calculate the expectation value of the operator $\bar{f} f$. While trying to interpret these quantities classically, we first need to normal order the product of the quantum fields:
\begin{align}\label{eq:calcFF}
     \langle:\bar{f} f:\rangle & \ = \ \langle \Psi|:\bar{f} f:|\Psi\rangle \  = \ \int \frac{d^3p_1}{(2\pi)^3}  \int \frac{d^3p_2}{(2\pi)^3}  \frac{1}{\sqrt{2E_{\textbf{p}_1}}} \frac{1}{\sqrt{2E_{\textbf{p}_2}}}\;  \nonumber\\ & \times \sum_{s,s'} \left[ \langle a^{s\dagger}_{\textbf{p}_1} a^{s'}_{\textbf{p}_2}\rangle \: \overline{u}^s(p_1) u^{s'}(p_2) \: e^{-i(\textbf{p}_1-\textbf{p}_2)\cdot \textbf{x}} +  \langle a^{s\dagger}_{\textbf{p}_1} b^{s'\dagger}_{\textbf{p}_2}\rangle \: \overline{u}^s(p_1) {v}^{s'}(p_2) \: e^{-i(\textbf{p}_1+\textbf{p}_2)\cdot \textbf{x}} \right.\nonumber\\
    & \left.  +\langle b^{s}_{\textbf{p}_1} a^{s'}_{\textbf{p}_2}\rangle \: \overline{v}^{s}(p_1) u^{s'}(p_2) \: e^{i(\textbf{p}_1+\textbf{p}_2)\cdot \textbf{x}} +  \langle b^{s'\dagger}_{\textbf{p}_2} b^{s}_{\textbf{p}_1}\rangle \: \overline{v}^s(p_1) v^{s'}(p_2) \: e^{i(\textbf{p}_1-\textbf{p}_2)\cdot \textbf{x}} 
    \right] \, ,
\end{align}
where we have used $\langle A\rangle = \langle \Psi|A|\Psi \rangle$ for brevity and the symbol :  : signifies normal ordering of the product. In Eq.~\eqref{eq:calcFF}, terms like $a^\dagger b^\dagger$ and $a\: b$ vanish, since they cannot be contracted because they act on different subspaces.  It is well known from quantum field theory at zero temperature that $a^{\dagger}a$ and $b^{\dagger}b$ are the number density operators for fermions and anti-fermions respectively. This can be generalized to finite temperature and density using the Fermi-Dirac distribution:
\begin{align}\label{eq:normND}
    \langle \Psi|a^{s\dagger}_{\textbf{p}_1} a^{s'}_{\textbf{p}_2}|\Psi\rangle \ & = \  n_f(\textbf{p}_1)\: \delta(\textbf{p}_1-\textbf{p}_2) \delta_{s,s'} \, , \\
\label{eq:normNDAF}
    \langle \Psi|b^{s\dagger}_{\textbf{p}_1} b^{s'}_{\textbf{p}_2}|\Psi\rangle \ & = \  n_{\bar f}(\textbf{p}_1)\: \delta(\textbf{p}_1-\textbf{p}_2) \delta_{s,s'} \, .
\end{align}
Eq.~\eqref{eq:normND} can be understood by integrating it over all momentum states which yields the total number density $N_f$:
\begin{equation}
  \int \frac{d^3p_1}{(2\pi)^3}\int \frac{d^3p_2}{(2\pi)^3}  \langle \Psi|a^{s\dagger}_{\textbf{p}_1} a^{s'}_{\textbf{p}_2}|\Psi\rangle \ = \ N_f~.
\end{equation}
Using the normalization of states $\overline{u}^s(p) u^{s}(p) = 2 m_f$, we obtain:
\begin{equation}\label{eq:ffMed}
   \langle \bar{f} f \rangle \ \equiv \ \langle \Psi|\bar{f} f|\Psi\rangle \ = \ g_f \int \frac{d^3p}{(2\pi)^3} \frac{m_f \:}{E_{\textbf{p}}}\left[n_f(\textbf{p})+n_{\bar f}(\textbf{p})\right] \, .
\end{equation}
Converting Eq.~\eqref{eq:ffMed} into an energy integral, we have:
\begin{align}
    \langle \bar{f} f \rangle  \ = \ \frac{g_f m_f}{2\pi}\int_{m_f}^\infty dk_0\:\sqrt{k_0^2-m_f^2} \left[n_f(k_0)+n_{\bar{f}}(k_0)\right] \, .
    \label{eq:ffMedE}
\end{align}
Notice that the integral form of Eq.~\eqref{eq:ffMedE} matches Eq.~\eqref{eq:sNSI} except for the pre-factors. This implies that generalizing the limiting cases for Eq.~\eqref{eq:ffMedE} is straightforward. 

\par Now to calculate $ \Delta m_{\nu, {\alpha\beta}}$ in Eq.~\eqref{eq:massC}, we need to solve Eq.~\eqref{eq:PhiEOM} for $\phi$. Considering $y_f \bar{f} f$ as a source term and neglecting the second term assuming low neutrino number density, we can write the solution as:
\begin{equation}
  \langle\phi\rangle(\textbf{x}) \ = \ -y_f \int d^3\textbf{x}^\prime \frac{\langle \bar{f} f \rangle(\textbf{x}^\prime)}{4\pi |\textbf{x}-\textbf{x}^\prime|} e^{-m_\phi(|\textbf{x}-\textbf{x}^\prime|)}~.
\end{equation}
Under assumptions of spherical symmetry of the medium, integrating over the angular variables yields the solution of the form:
\begin{equation}\label{eq:NSILongR}
    \Delta m_{\nu,{\alpha\beta}}(r) \ = \   \frac{y_f y_{\alpha\beta}}{m_\phi \: r}\left( e^{-m_\phi r}\int_0^r x  \: \langle \bar{f} f \rangle \: \sinh{(m_\phi\: x)}\: dx + \sinh{(m_\phi \: r)} \int_r^\infty x  \: \langle \bar{f} f \rangle \: e^{-m_\phi\: x} \: dx  \right) ~.
\end{equation}
We have worked out Eq.~\eqref{eq:NSILongR} in the relativistic limit for two different density profile distributions in Appendix \ref{appendix:4}. While we do not use these analytic results in our numerical analysis, these special cases can give insight for general situations.  We use actual density profiles of the Sun and supernovae in our numerical calculations, integrating the relevant integrals exactly.

\section{Experimental constraints on couplings} \label{sec:4}

In this section we explore two specific scenarios: 
\begin{enumerate}
    \item [(i)] scalar $\phi$ coupling only to electrons and neutrinos, and
    \item [(ii)] scalar $\phi$ coupling only to nucleons and neutrinos. 
\end{enumerate}
Here neutrinos can be either Dirac or Majorana in nature. In this section, we discuss experimental constraints on the couplings and mass of $\phi$ in the aforementioned scenarios.

\par In accordance to Eq.~\eqref{Dirac}, the scalar coupling to electron is denoted by $y_e$. On the other hand, the scalar coupling to quark cannot be probed directly but only measurable through their effect with scalar-nucleon interaction. Thus, we present the experimental constraints on  scalar-nucleon coupling labeled as $y_N$. The conversion from quark level couplings $y_q$ to $y_N$ is discussed later in Sec.~\ref{sec:MesonD}.

\subsection{Constraints on $y_e$ and $y_N$}
\subsubsection{Anomalous electron magnetic moment}

A scalar coupling with the electrons will contribute to the electron anomalous magnetic moment $(g-2)_e$ given by \cite{Liu:2018xkx}: 
\begin{equation}
  \Delta a_e \ = \ \frac{1}{8\pi^2}\int_0^1 dx \frac{(1-x)^2(1+x)y_e^2}{(1-x)^2+x(m_\phi/m_e)^2}~.
\end{equation}
There is currently a $2.4 \sigma$  discrepancy between the experimentally inferred value and SM prediction for  $\Delta a_e= a_e^{\rm exp}-a_e^{\rm SM}=
(-88\pm36) \times 10^{-14}$~\cite{Tanabashi:2018oca}. A light scalar can potentially make this discrepancy worse, as it gives a positive contribution, and thus provides a useful limit on scalar NSI parameters.  Using the $3\sigma$ value for the $\Delta a_e$, the allowed region in the  $y_e-m_\phi$ plane is obtained. This constraint is shown in Figs.~\ref{fig:YeD} and \ref{fig:YeM}, labeled as $(g-2)_e$.
This constraint yields an almost constant upper bound of $y_e < 3.4 \times 10^{-6}$ for light scalar mediators. 

\subsubsection{Fifth force experiments}

These experiments measure the presence of fifth forces as deviation from the Newtonian gravitational potential between a given source mass and a test mass, which is parametrized as follows:
\begin{equation}
    V(r) \ = \ -\frac{G m_1 m_2}{r}\left(1+ \alpha e^{-r/\lambda} \right)~.
    \label{eq:fifth}
\end{equation}
Given an interaction vertex of the form $y_f \bar{f}\phi f$ as in Eqs.~\eqref{Dirac} and \eqref{Majorana},
consider the scattering of two distinguishable fermions in the non-relativistic limit. The corresponding Yukawa potential for the interaction is given by (see Sec.~4.7 of Ref.~\cite{Peskin:1995ev}):
\begin{equation}
    V(r) \ = \ -\frac{y_f^2}{4 \pi r}\: e^{-m_\phi r} \, ,
    \label{eq:fifth2}
\end{equation}
where $r$ is the distance between the scattering particles.
\par For experiments detailed in Refs.~\cite{Irvine85,Colorado02,EotWash07,Stanford08,HUST12,HUST+16} in the range $\lambda= 10^{-6} \text{ to } 10^{2}$ m, the constraints provided on $\alpha$ in Eq.~\eqref{eq:fifth} are not directly applicable to $y_f$ in Eq.~\eqref{eq:fifth2}. Therefore, we will translate the constraints on $\alpha$ to those on $y_f$ for our case. Assuming a particle (e.g. lepton, quark) couples to the scalar mediator with strength $q$ and each interacting body contains $N$ number of these particles, the potential between two extended bodies can be written as:
\begin{equation}
    V_\phi (r) \ = \ -\frac{N_1q_1 \: N_2q_2}{4 \pi r} \: e^{-m_\phi r}~.
\end{equation}
We identify the inverse of the length scale $\lambda$ as mass of the scalar particle $\phi$. Thus, we have: 
\begin{equation}
    \alpha \ = \  \frac{N_1q_1 \: N_2q_2}{4 \pi G m_1 m_2} \ = \ \frac{q_1 \: q_2}{4 \pi G A_1 A_2 u^2}
     \ = \ \frac{1}{4 \pi G u^2}\frac{q_1}{A_1}\frac{q_2}{A_2} \, ,
\end{equation}
where we have used the relation $m= N A u$ ($A$= mass number, $u$ = 1 atomic mass unit) and $G$ is the gravitational constant. For bounds on $y_e$, the coupling strength will be proportional to the lepton number ($L$), which is identical to atomic number ($Z$) for a given material, i.e., $q= Zy_e $, leading to
\begin{equation}
    \alpha \ = \ \frac{y_e^2}{4 \pi G u^2}\frac{Z_1}{A_1}\frac{Z_2}{A_2}~.
    \label{eq:52}
\end{equation}

Values for charge to mass number ratio for test and source masses can be obtained from the experimental setups as given in Table \ref{table:RatioZA}. These are shown in Figs.~\ref{fig:YeD} and \ref{fig:YeM} by the labels I to VII. Similar results follow for coupling to the nucleons $y_N$ by replacing the atomic numbers ($Z$) by mass numbers ($A$) in Eq.~\eqref{eq:52}. This implies that constraints on $y_N$ will be independent of the material used in the experiment. These limits are shown in Figs.~\ref{fig:YnD} and \ref{fig:YnM}.
\begin{table}[t!]
\centering
\begin{tabular}{ |c|c|c|c|c| } 
\hline
Label & References & Source Mass Composition & Test Mass Composition & $\frac{Z_1}{A_1}\frac{Z_2}{A_2}$ \\
\hline
I & Stanford \cite{Stanford08} & Gold, Silicon & Gold & 0.1804 \\
II & Colorado \cite{Colorado02} & Tungsten & Tungsten & 0.1621 \\
III & Eot-Wash'07 \cite{EotWash07} & Molybdenum, Tantalum & Molybdenum & 0.1839 \\
IV & HUST'12 \cite{HUST12} & Tungsten & Tungsten & 0.1621 \\
V & HUST+ '16 \cite{HUST+16} & Tungsten & Tungsten & 0.1621 \\
VI & Irvine A \cite{Irvine85}& Copper & Copper & 0.2159 \\
VII & Irvine B \cite{Irvine85}& Stainless Steel & Copper & 0.2116 \\
\hline
\end{tabular}
\caption{The compositions of source and test masses used in the experiment listed and the corresponding values of ratio $\frac{Z_1}{A_1}\frac{Z_2}{A_2}$.}
\label{table:RatioZA}
\end{table}

\par Additional constraints on $y_e$ and $y_N$ can be directly obtained from Ref.~\cite{ADELBERGER2009102} which used  experiments in the range $\lambda= 10^{-1} \text{ to } 10^{13}$ m and the corresponding limits on 
\begin{equation}
   \widetilde{\alpha} \ = \ \frac{y_{e(N)}^2}{4 \pi G u^2} \, .
\end{equation}
This constraint is labeled as ``Torsional Balances" in Fig.~\ref{fig:YeD}, \ref{fig:YnD}, \ref{fig:YeM}, and  \ref{fig:YnM}. It can be seen from these figures that fifth-force experiments constrain both couplings $y_e$ and $y_N$ with an upper bound in the range $10^{-25} -10^{-15}$ for $m_\phi < 0.1$ eV.

\subsubsection{Constraints from Stellar and Supernova Cooling}\label{sec:RGHBSN}

\textbf{\underline{$\phi-e$ coupling}:} The production of the light scalar $\phi$ in stellar bodies can lead to a new channel for energy loss leading to rapid cooling. This can help severely constrain the interaction of a scalar with electrons. The dominant production of this scalar is via its resonant mixing with the longitudinal component of the photon in the plasma~\cite{Hardy:2016kme}. The extra energy loss processes in red giants (RG) can delay their onset of helium ignition and can change the helium-burning lifetime of the horizontal branch (HB) stars, in disagreement with the stellar models that match observations. For bounds from supernova, the energy loss from production of a scalar is required to be less than that of SN1987A neutrino burst. The energy loss rate from resonant production of a scalar with a plasmon is given by \cite{Hardy:2016kme,Knapen:2017xzo}
\begin{equation}
    Q_{\rm res} \ \simeq \ \frac{\omega_L}{4\pi}\left( \frac{\omega_L}{m_\phi}\Pi^{\phi L} \right)\frac{1}{e^{\frac{\omega_L}{T}}-1} \, ,
\end{equation}
where $\omega_L$ is the resonant frequency and $\Pi^{\phi L}$ is the mixing of the scalar with the longitudinal component of the photon in the medium, given by 
\begin{equation}
   \Pi^{\phi L} \ \simeq \ \frac{y_e e m_e^{\rm eff}m_\phi}{\pi^2 k} \int_0^\infty dp \: v^2 \left[n_e(E_p)+n_{\overline{e}}(E_p)\right]\left[ \frac{\omega_L}{v_k}\log\left(\frac{\omega_L + vk}{\omega_L - vk} \right)- \frac{2 m_\phi^2}{\omega_L^2 -k^2v^2}\right] \, , 
\end{equation}
where $v=p/E_p$ is the electron velocity, $m_e^{\rm eff}$ is the effective thermal mass of the electron and $k=\sqrt{\omega_L^2-m_\phi^2}$ is the 3-momentum of the scalar mediator $\phi$, where $E_\phi=\omega_L$ due to the resonant production of scalar. Ref.~\cite{Knapen:2017xzo} considers the resonant production process as dominant over the Compton scattering or electron-ion interactions. 

\par For large values of the coupling, the scalar can get trapped inside the star/supernova. This capture would help alleviate the stringent upper bound on the coupling $y_e$. To derive the trapping limit, the detailed balance of production and absorption rates is used, i.e., 
\begin{align}
\Gamma_{\rm prod}(E_\phi) \ = \ e^{-\frac{E_\phi}{T}}\Gamma_{\rm abs}(E_\phi) \, .
\end{align}
Since we are only interested in ultra-light mediators with $m_\phi<1 \text{ MeV}$, the absorption through the decay channel $\phi \rightarrow e^+ e^-$ is absent for our purposes. Thus, the absorption rate from the resonant mixing yields a mean free path length $\lambda$ given by:
\begin{equation}
    \lambda \ = \ \frac{1}{\Gamma_{\rm abs}(E_\phi)} \ \sim \ \frac{E_\phi^4}{Q_{\rm res}}~.
\end{equation}
By requiring the mean free path to be shorter than $R= 10 \text{ km}$, which is the typical size of a supernova core, we derive a bound on the coupling $y_e$, as shown in Figs.~\ref{fig:YeD} and \ref{fig:YeM}, labeled ``SN1987A". 

\par In case of SN1987A, constraints on $y_e$ range from $10^{-9}$ to $10^{-7}$ for scalar mediators lighter than the electron. Even stronger constraints are obtained from HB/RG stars with an upper bound of $y_e \sim 10^{-15}$ for light scalar mediators.\\
\newline
\textbf{\underline{$\phi-N$ coupling} :} The constraints are similar to the $\phi-e$ coupling case. In HB and RG stars with typical temperatures of $10$ keV, the main constraints for scalar coupling to nucleon in the literature are derived using Compton scattering, $\gamma + \text{He} \rightarrow \text{He} + \phi$, as the dominant process. It is required that the new energy loss per unit mass should  be less than $\epsilon<10$ erg/g/s \cite{Raffelt:1996wa}. As shown in Ref.~\cite{Hardy:2016kme}, resonant production through $\phi$ mixing with a photon can increase the energy loss for low scalar masses and therefore the $\phi$ coupling to nucleon is highly constrained.

\par  The constraints from a supernova comes from scalar production through bremsstrahlung process $N + N \rightarrow N + N + \phi$ \cite{Ishizuka:1989ts}. Bounds on the coupling can be obtained by requiring the energy loss to be less than the energy contained in the neutrino burst, i.e., $\epsilon<10^{19}$ erg/g/s~\cite{Raffelt:1996wa}. Similarly, the trapping regime of the scalar being reabsorbed can be derived using the detailed balance between the absorption and production rates. Requiring the mean free path $\lambda \propto \epsilon \rho / T^4$ to be smaller than 10 km yields the constraint on $y_N$ \cite{Knapen:2017xzo}, as shown in Figs.~\ref{fig:YnD} and  \ref{fig:YnM}.

\par In case of SN1987A, constraints on $y_N$ range from $10^{-10}$ to $10^{-7}$ for scalar mediators lighter than electron. Similar to $y_e$, stronger constraints are obtained from HB/RG stars with an upper bound of $y_N \sim  10^{-12}$ for light scalar mediators.

\subsubsection{Meson decays}\label{sec:MesonD}
A light scalar coupling to nucleons can be produced in meson decays. The only process of interest in this case is a charged Kaon decay to a charged pion and the scalar: $K^+ \rightarrow \pi^+ \phi$. This production cross section is highly constrained from the measurement of branching ratios from charged Kaon decay : $\text{Br }(K^+ \rightarrow \pi^+ \overline{\nu}\nu) < 1.7 \times 10^{-10} $ \cite{Tanabashi:2018oca}. 

\par Using the low-energy effective Lagrangian formalism presented in Ref.~\cite{Batell:2018fqo}, the branching ratio for the process in consideration is given by
\begin{equation}
    {\rm BR}(K^+ \rightarrow \pi^+ \: \phi) \ = \ \frac{(3y_u G_F f_\pi f_K B)^2}{32\pi m_{K^+}\Gamma_{K^+}}|V_{ud}V_{us}|^2\:\lambda^{1/2}\left(1,\frac{m_\phi^2}{m_{K^+}^2},\frac{m_{\pi^+}^2}{m_{K^+}^2}\right) \, ,
\end{equation}
where $B=\frac{m_{\pi}^2}{m_u+m_d}$ and  $\lambda(a,b,c)=a^2+b^2+c^2-2ab-2bc-2ac$. 
Matching the nucleon level interaction to the effective Lagrangian:
\begin{equation}
    \mathcal{L} \ \supset \ y_{N}\overline{N}N\phi \, ,
\end{equation}
where $N= p,\, n$, the nucleon coupling $y_N$ can be written in terms of fundamental quark -level couplings $y_u(y_d)$ as :
\begin{equation}
    y_{N} \ = \ \sum_{q} y_q g^q_S \, ,
\end{equation}
where $g^q_S$ is the nucleon scalar charge. We assume that the scalar couples equally to the up and down quark i.e. $y_u = y_d$. The effective nucleon couplings to a scalar is then given by 
\begin{align}\label{eq:Q2N}
    y_{N} & \ = \ y_u \left(g^u_S + g^d_S \right) \simeq 9.47\: y_u \, ,
\end{align}
where we have used $g^u_S=5.20$ and $g^d_S=4.27$ \cite{Alexandrou:2017qyt}. This constraint is labeled as ``$K^+ \rightarrow \pi^+ \phi$" in Figs.~ \ref{fig:YnD} and \ref{fig:YnM}. It yields an almost constant upper bound of $y_N \sim 2.3 \times 10^{-5}$ for light scalar mediators.

\subsubsection{Big Bang Nucleosynthesis}
\textbf{\underline{$\phi-e$ coupling}:} In early Universe, the scalar mediator $\phi$ can be in thermal equilibrium with the SM particles through annihilation ($e^+e^-\rightarrow\gamma\phi$) and Compton scattering ($e^- \gamma \rightarrow e^- \phi$). In the limit $s\gg m_\phi^2, m_e^2  $,  the cross sections for these processes are \cite{Knapen:2017xzo}
\begin{align}
    \sigma_{e \gamma \rightarrow e \phi} \ & \approx \  \frac{\alpha_e y_e^2}{s}\left[\log\left(\frac{s}{m_e^2 + m_\phi^2} \right)+\frac{5}{2} \right] \, ,\\
    \sigma_{e e \rightarrow \gamma \phi}\ & \approx \  \frac{2\alpha_e y_e^2}{s}\log\left(\frac{s}{4m_e^2} \right)~,
\end{align}
where $\alpha_e\equiv e^2/4\pi$ is the fine-structure constant. 
The thermally averaged cross section for these two processes are given below:
\begin{align}
   \langle \sigma_{e \gamma \rightarrow e \phi}\:v \rangle \ & = \  \frac{1}{16m_e^2T^3K_2(m_e/T)}\int_{m_e^2}^\infty ds \: \sigma(s-m_e^2)\sqrt{s}K_1\left(\frac{\sqrt{s}}{T}\right) \, ,\\
    \langle \sigma_{e e \rightarrow \gamma \phi}\:v \rangle \ & = \  \frac{1}{8m_e^4T(K_2(m_e/T))^2}\int_{4m_e^2}^\infty ds \: \sigma(s-4m_e^2)\sqrt{s}K_1\left(\frac{\sqrt{s}}{T}\right)~.
\end{align}
If $\phi$ enters equilibrium with electrons before $T \sim 1\text{ MeV}$, it can decrease the deuterium abundance which is in conflict with observations \cite{Knapen:2017xzo}. In our case, the mediator thermalizes if the thermally averaged cross section exceeds the Hubble expansion rate $H(T)\sim 1.66 \sqrt{g_*}T^2/M_{\text{Pl}}$ (where $g_*$ is the number of relativistic degrees of freedom and $M_{\rm Pl}$ is the Planck mass) at $T=1 \text{ MeV}$. This yields an upper bound of $y_e = 5 \times 10^{-10}$ for ultra-light scalar mediators, independent of $m_\phi$.

Note that LEP measurements of the  Bhabha scattering cross-section ($e^+e^-\rightarrow e^+e^-$) can also constrain the coupling $y_e$ through $s$ and $t$-channel $\phi$ exchange, but we estimate it to be only at  $\mathcal{O}(0.1)$ level~\cite{ALEPH:2004aa, Babu:2019mfe}.
\\ \\
\textbf{\underline{$\phi-N$ coupling} :} In this case, we require that the scalar $\phi$ thermalizes around the QCD phase transition temperature. This will help dilute the relativistic degrees of freedom ($N_{\rm eff}$) until the nucleosynthesis phase is reached. Otherwise, the scalar $\phi$ will be in equilibrium with SM and will have a significant contribution to relativistic degrees of freedom ($\Delta N_{\rm eff}=4/7$) at the time of BBN, in tension with the current measurements from Planck~\cite{Aghanim:2018eyx}. Thus, we require that the interaction rate should be lower than the Hubble rate at $T=200$ MeV. We can estimate the rate of $\phi$ production from the processes like $u\Bar{u}\rightarrow\phi$ (and $d\Bar{d}\rightarrow\phi$) as $\Gamma_\phi \sim y_u^2T$. This should be compared with Hubble rate $H(T)\sim 1.66 \sqrt{g_*}T^2/M_{\text{Pl}}$. This condition leads to a stringent constraint on $y_u< 2.63\times 10^{-10}$. Converting the quark-scalar coupling to nucleon level coupling using Eq.~\eqref{eq:Q2N} , we get $y_N< 2.49 \times 10^{-9}$.

\subsection{Experimental Constraints on $y_\nu$} \label{sec:4.2}
\textbf{\underline{Dirac $\nu-\phi$ coupling}:} 
The analysis in this case is similar as for the $\phi-N$ coupling. If the scalar $\phi$  thermalizes (even partially) in the early Universe, it introduces additional degrees of freedom that contribute to the total entropy~\cite{Escudero:2019gfk}. We require that the scalar $\phi$, as well as the right-handed neutrinos, should decouple from the thermal plasma at a temperature above the QCD phase transition temperature which will dilute the  $\Delta N_{\rm eff} = 3+\frac{4}{7} \sim 3.57$ by the time BBN occurs, in agreement with the currently allowed range from Planck~\cite{Aghanim:2018eyx}. Thus, requiring that the interaction rate of processes like $\nu\Bar{\nu}\rightarrow\phi$ should be lower than the Hubble rate at $T=200$ MeV yields an upper bound of $y_\nu\sim 2.6\times 10^{-10}$.
\\ \\
\textbf{\underline{Majorana $\nu-\phi$ coupling}:} 
Presence of NSI can lead to re-thermalization of the neutrinos, which otherwise decouple at $T\sim 1$ MeV in the standard scenario. This can leave a signature in the cosmological observables. The analysis in Ref.~\cite{Forastieri:2019cuf}  constrains the couplings in the secret interaction of neutrinos with a light mediator. Assuming model independence, we use the upper bound on coupling $y_\nu$ from Ref.~\cite{Forastieri:2019cuf}, which yields a stringent limit of $y_\nu < 2 \times 10^{-7}$. 

The next-generation CMB experiments, such as CMB-S4~\cite{Abazajian:2016yjj} which will have better sensitivity to departures from the $\Lambda$CDM paradigm could test such neutrino self-interactions mediated by light scalars, as discussed here. 

Additional constraints on $y_\nu$ exist from neutrino self-interactions within astrophysical sources like core-collapse supernovae~\cite{Shalgar:2019rqe} with high  neutrino number densities of $n_\nu\sim {\cal O}(10^{38})~{\rm cm}^{-3}$, where they can lose energy via higher-order processes like $2\nu\to 4\nu$ and may be unable to transfer enough energy to the stalled supernova shock wave to revive it, halting the explosion altogether~\cite{Bethe:1984ux, Shalgar:2019rqe}. Similarly, elastic scattering of astrophysical neutrinos off the cosmic neutrino background as they propagate to Earth would distort the energy spectrum of the astrophysical neutrinos by introducing a deficit at high energies and a pileup at low energies, potentially falling below the energy threshold for detection, as well as delaying their arrival time on Earth, compared to their electromagnetic-wave counterpart~\cite{Kolb:1987qy, Shalgar:2019rqe}. However, these astrophysical constraints on $y_\nu$ turn out to be  much weaker than the cosmological constraints discussed above for light scalars with $m_\phi\lesssim 1$ MeV.

It should also be pointed out that there are other weaker constraints applicable in our scenario but not relevant to the scalar NSI discussion here. For example, coherent elastic neutrino-nucleus scattering data by COHERENT experiment constrains  $y_N$ only at the $\mathcal{O}(1)$ level for the values of the $y_\nu$ used in this work~\cite{Farzan:2018gtr}.

\section{Thermal mass of scalar $\phi$}\label{sec:TM}
\begin{figure}[t]
    \centering
    \includegraphics[scale=1.3]{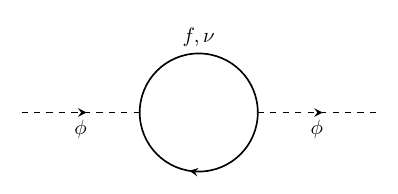}
    \caption{Feynman diagram responsible for the thermal mass of the scalar $\phi$.}
    \label{fig:D3}
\end{figure}

If the interactions of the scalar $\phi$ with the medium are significant enough, then it might get thermalized with the medium. Since the scalar field in consideration is ultra-light, medium effects might lead to substantial correction to the vacuum mass of the $\phi$. The medium induced mass at one-loop is shown in the Feynman diagram in Fig.~\ref{fig:D3}.  The relevant contribution to the mass of $\phi$  at finite density and temperature is given by:
\begin{equation}
    \mathcal{M} \ = \ 4 y_f^2 \int \frac{d^4k}{(2\pi)^4} \left(k^2-\frac{p^2}{4}+m_f^2\right)\left[ \frac{\Gamma(k+p/2)}{(k-p/2)^2-m_f^2} + \frac{\Gamma(k-p/2)}{(k+p/2)^2-m_f^2} \right]
    \label{eq:ThermalMassPhi}~.
\end{equation}
We refer the reader  to Appendix \ref{appendix:2} for the evaluation of the scalar mass integral. In the limit $m_\phi \rightarrow 0$, the mass correction for scalar is found to be:
\begin{align}
    \Delta m_\phi^2 \ = \ \frac{y_f^2}{\pi^2} \int_{m_f}^\infty dk_0\: n_f(k_0) \sqrt{k_0^2-m_f^2}   \, .
   \label{eq:ThermalPhi}
\end{align}
Note that the same integral appears in Eq.~\eqref{eq:sNSI}. Thus, using the known limiting forms of the integral (cf.~Appendix~\ref{appendix:1}), we obtain: 
\begin{numcases}{ \Delta m_\phi^2 = \ }
\frac{y_f^2}{m_f}\left(N_f+N_{\bar f}\right) & \textbf{     ($\mu, T\ll m_f$)} \label{eq:TP1} \\
 \frac{y_f^2}{2}\left(\frac{3}{\pi}\right)^\frac{2}{3} \left(N_f^{2/3}+N_{\bar f}^{2/3}\right) & \textbf{($\mu>m_f\gg T$)} \label{eq:TP2} \\
 \frac{y_f^3 }{3 }\left(\frac{\pi^2}{12 \:\zeta(3)}\right)^\frac{2}{3} \left(N_f^{2/3}+N_{\bar f}^{2/3}\right) & \textbf{ ($\mu<m_f\ll T$)}~. \label{eq:TP3}
\end{numcases}
These expressions are also applicable to Majoron ($J$) propagation in a medium with pseudoscalar interactions of the form $\bar{\nu}\gamma^5 J \nu$. For example, in the early Universe, Majoron  propagating in a neutrino background will have a mass given by the high-temperature limit, which will be approximately $m_J\simeq y_\nu T$ [cf.~Eqs.~\eqref{eq:TP3} and \eqref{eq:A14}].

Eq.~(\ref{eq:TP3}) will also be relevant  to deriving neutrino self-interaction limits from early Universe cosmology. CMB anisotropies strongly depend on the anisotropy of the neutrino field. Neutrino self-interactions would isotropize the neutrino field, affecting the CMB. It has been found that CMB anisotropy data constrain such interactions to be $(y_\nu^2/m_\phi^2) \leq (3~{\rm MeV})^{-2}$ (for $m_\phi > 1$ keV) \cite{Kreisch:2019yzn}.  If the scalar field indeed thermalizes with the medium, which occurs for $y_\nu \geq 10^{-10}$ or so, then one should use the thermal mass of $\phi$, Eq.~(\ref{eq:TP3}) in this constraint, which can weaken the constraint significantly.  In cosmological simulations involving a light scalar, the thermal mass effects of Eq.~(\ref{eq:TP3}) should be included.  Such interactions may be testable in future CMB and large-scale structure observations through the thermally induced mass in such settings.

In the limit when $m_\phi \rightarrow 0$ but acquires a thermal mass, the scalar NSI expression Eq.~\eqref{eq:sNSI} takes a special form:
\begin{equation}
     \Delta m_{\nu,{\alpha\beta}} \ = \ 
\frac{ y_{\alpha\beta}}{y_f}m_f~.
\label{eq:thSNSI}
\end{equation}
Note that Eq.~\eqref{eq:thSNSI} is independent of the scalar mass $m_\phi$ in this limit. This scenario may be realized in supernovae, provided that $\phi$ has significant interactions with matter. From discussions in Sec.~\ref{sec:RGHBSN}, it is clear that for high enough values of $y_e$ or $y_N$, the scalar gets trapped and thermal correction to the mass should be taken into account. Thus, in case of thermalization of the scalar, Eq.~\eqref{eq:thSNSI} should be used in lieu of Eqs.~\eqref{eq:c1}, \eqref{eq:c2} and \eqref{eq:c3}.

\section{Quantum-mechanical bound on light scalar mass}\label{sec:QM}

Here we show that the uncertainty principle of quantum mechanics sets a lower limit on the minimum $q^2$ that appears in neutrino forward scattering.  This limit applies to a neutrino propagating through Earth, where it interacts either with electrons in atoms, or with nucleons inside the nuclei.   

Consider $\nu_\alpha-e$ elastic scattering.  Working in the rest frame of the electron, the initial and final four-momenta of the electron can be written as
\begin{equation}
p^\mu \ = \ (m_e, \,0,\,0,\,0)\, , \qquad p'^{ \mu} \ = \  (\sqrt{p_e^2+m_e^2},\, 0,\, 0,\, p_e) \, ,
\end{equation}
where $p_e$ is the recoil momentum of the electron.  The $q^2$ related to coherent forward scattering is then 
\begin{equation}
q^2 \ = \ (p'-p)^2 \ = \ 2 m_e (m_e-\sqrt{p_e^2+m_e^2}) \ \simeq \  -p_e^2, 
\end{equation}
where in the second step $q^2 \ll m_e^2$ is assumed.  

Now, the recoil momentum of the electron is subject to the uncertainty relation.  Its position is not precisely known inside the atom, so we have
\begin{equation}
\Delta p\, \Delta x \ \gtrsim \ \hbar~.
\end{equation}
When we set $q^2 = 0$ in the computation of forward scattering, we only know this up to an uncertainty in $q^2$ given by (setting $\hbar=1$)
\begin{equation}
    q^2 \ \simeq \ p_e^2 \ \sim \ \frac{1}{(\Delta x)^2}~.
\end{equation}
Using $\Delta x = 140 \times 10^{-8}$ cm, which is the radius of $^{26}$Fe  -- the most abundant element in Earth's matter, one obtains for the uncertainty in $q^2$ to be
\begin{equation}
q^2_{y_e} \ \approx \ (14 ~{\rm eV})^2~.
\end{equation}
Thus, when the mediator mass becomes much smaller than $14$ eV, one should use this quantum mechanical cut-off in computing scalar NSI. Similarly for coupling to nucleon, the cut-off would be given by the inverse of the nuclear radius of $^{26}$Fe. Using nuclear diameter $\Delta x = 9.6$ fm, we obtain 
\begin{equation}
q^2_{y_N} \ \approx \ (21 ~{\rm MeV})^2~.
\end{equation}

These rough quantum-mechanical bounds can be better motivated by using atomic/nuclear form factors for coherent forward scattering. In Earth, the expression for scalar NSI will get modified with the inclusion of a form factor. 
\begin{equation}\label{eq:EnSI}
    \Delta m_{\nu, {\alpha\beta}} \ = \ 
\frac{y_f y_{\alpha\beta}N_f}{m_\phi^2-q^2}F(m_\phi^2) \, ,
\end{equation}
The original result in Eq.~\eqref{eq:c1} was obtained by setting $q^2=0$ and $F(m_\phi^2)=1$, but if the mass of the scalar $m_\phi\rightarrow0$ then the denominator is not well-defined. This is remedied by the atomic form factor $F(m_\phi^2)$ which is of the form \cite{hubbell1979relativistic}:
\begin{equation}
    F(m^2) \ = \ \frac{m^2}{m^2+q_0^2} \, ,
\end{equation}
where $q_0=1/4\pi a_0$ and $a_0$ is the radius of the first orbit for hydrogen-like atoms. Similar qualitative results should apply for the outermost-orbit electrons in $^{26}$Fe. For high values of $m^2_\phi\gg q_0^2$, $F(m_\phi^2)\sim 1$ as expected. Thus, the vanishing $q^2$ limit is well-defined and yields the original result in Eq.~\eqref{eq:c1}. Difference appears in the regime  $m_\phi^2\ll q_0^2$, where $F(m_\phi^2)\sim m_\phi^2/q_0^2$. The form of Eq.~\eqref{eq:EnSI} in the low $m_\phi$ limit and with $q^2\to 0$ is thus given by:
\begin{equation}
    \Delta m_{\nu,{\alpha\beta}} \ = \ 
 \frac{y_f y_{\alpha\beta}N_f}{q_0^2}~,
\end{equation}
which is independent of $m_\phi$. This result agrees with the quantum-mechanical bound discussed above based on the uncertainty principle. 

\par When a scalar mediator couples to the electron, from fifth force constraints either the mass of the mediator should be larger than a keV, or its coupling to the electron should be extremely weak, of order $10^{-24}$.  For such tiny couplings, to generate scalar NSI in the observable range, one could naively make the mediator mass of order $10^{-8}$ eV.  In this case, the quantum-mechanical intrinsic bound should be applied for computing forward scattering. The result is that scalar NSI arising from coupling to electrons cannot be in the observable range for neutrino propagation in Earth. 

These quantum-mechanical limits are not applicable to Sun or supernovae due to the absence of bound states in them. The major baryonic component in Sun and supernovae is ionized hydrogen gas (protons) and neutrons respectively. Thus, the neutrinos scatter off against either free electrons or the protons/neutrons inside these stellar bodies. For the relevant neutrino energies of ${\cal O}({\rm keV}-{\rm MeV})$, the protons/neutrons behave as point particles, and therefore, the finite-size effect discussed above is not applicable to them.


\section{Numerical results }\label{sec:discussion}
\begin{figure}[htbp]
  \centering
  \includegraphics[width=0.98\textwidth]{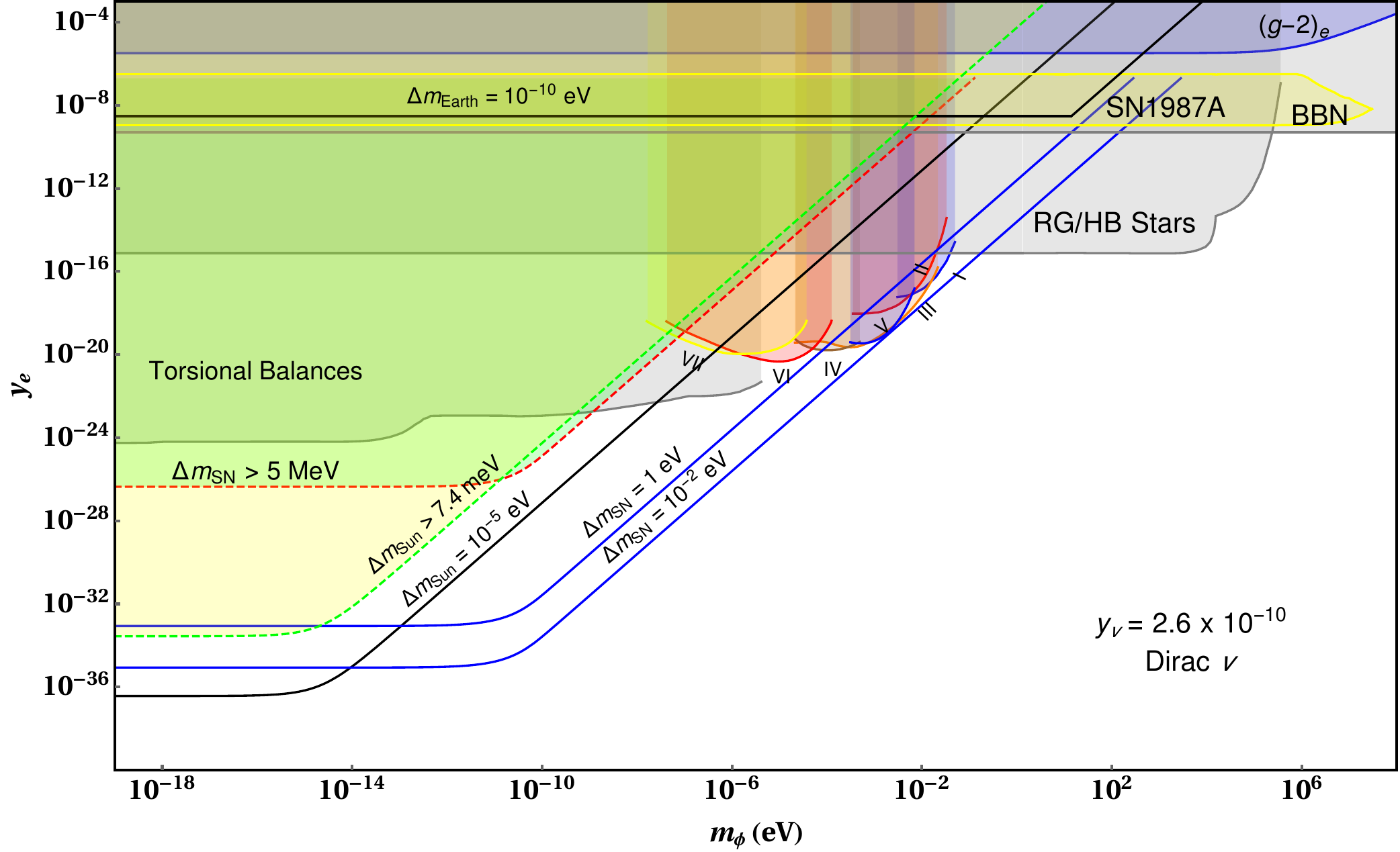}
    \caption{Different experimental constraints on Yukawa coupling of scalar to electron for the case of Dirac neutrinos. The shaded regions are excluded. Some representative values of scalar NSI in Earth, Sun and supernova are also shown.}
    \label{fig:YeD}

  \vspace*{\floatsep}

  \includegraphics[width=0.98\textwidth]{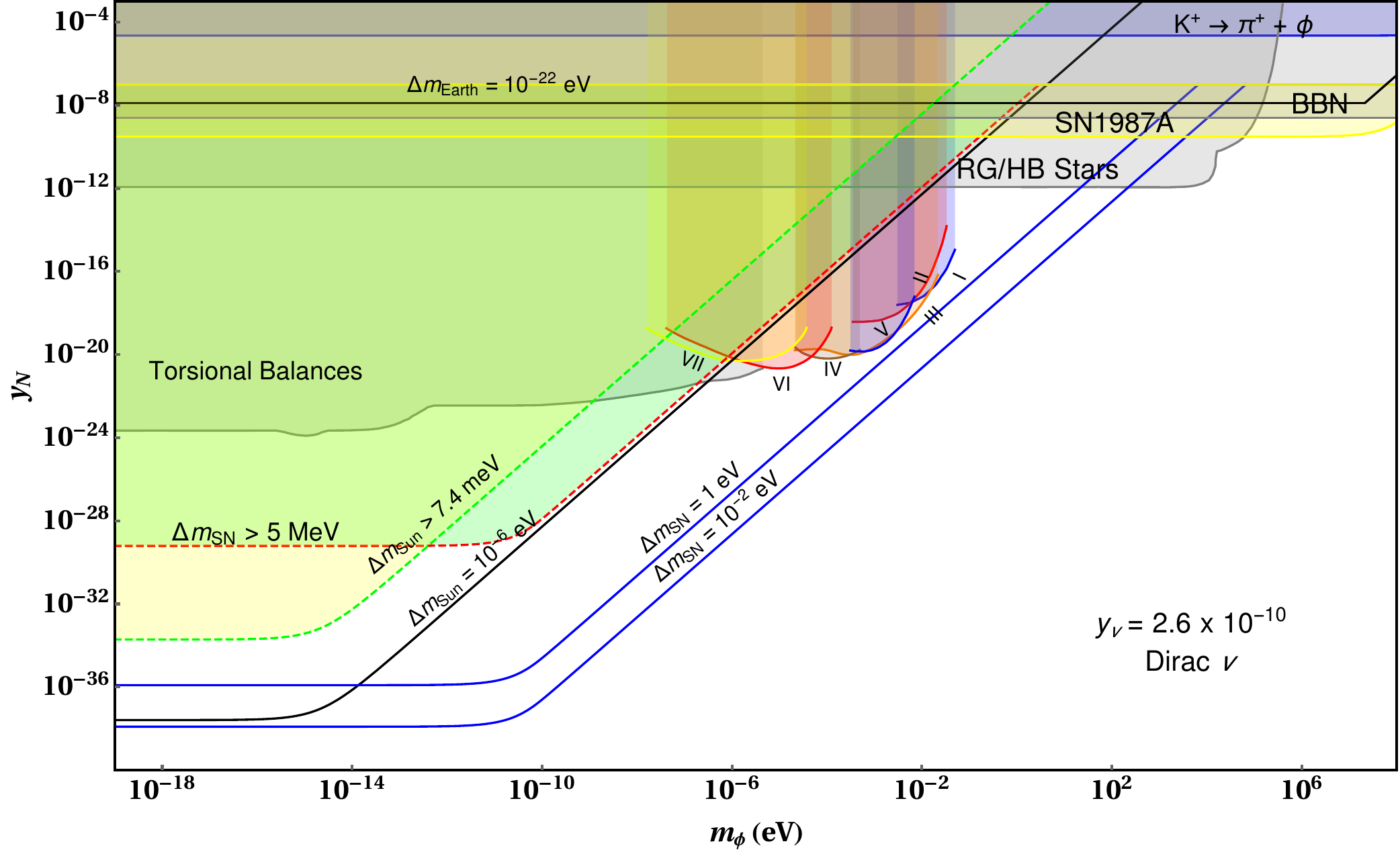}
    \caption{Same as in Fig.~\ref{fig:YeD}, but for scalar coupling to nucleons.}
    \label{fig:YnD}
\end{figure}

\begin{figure}[htbp]
  \centering
    \includegraphics[width=0.98\textwidth]{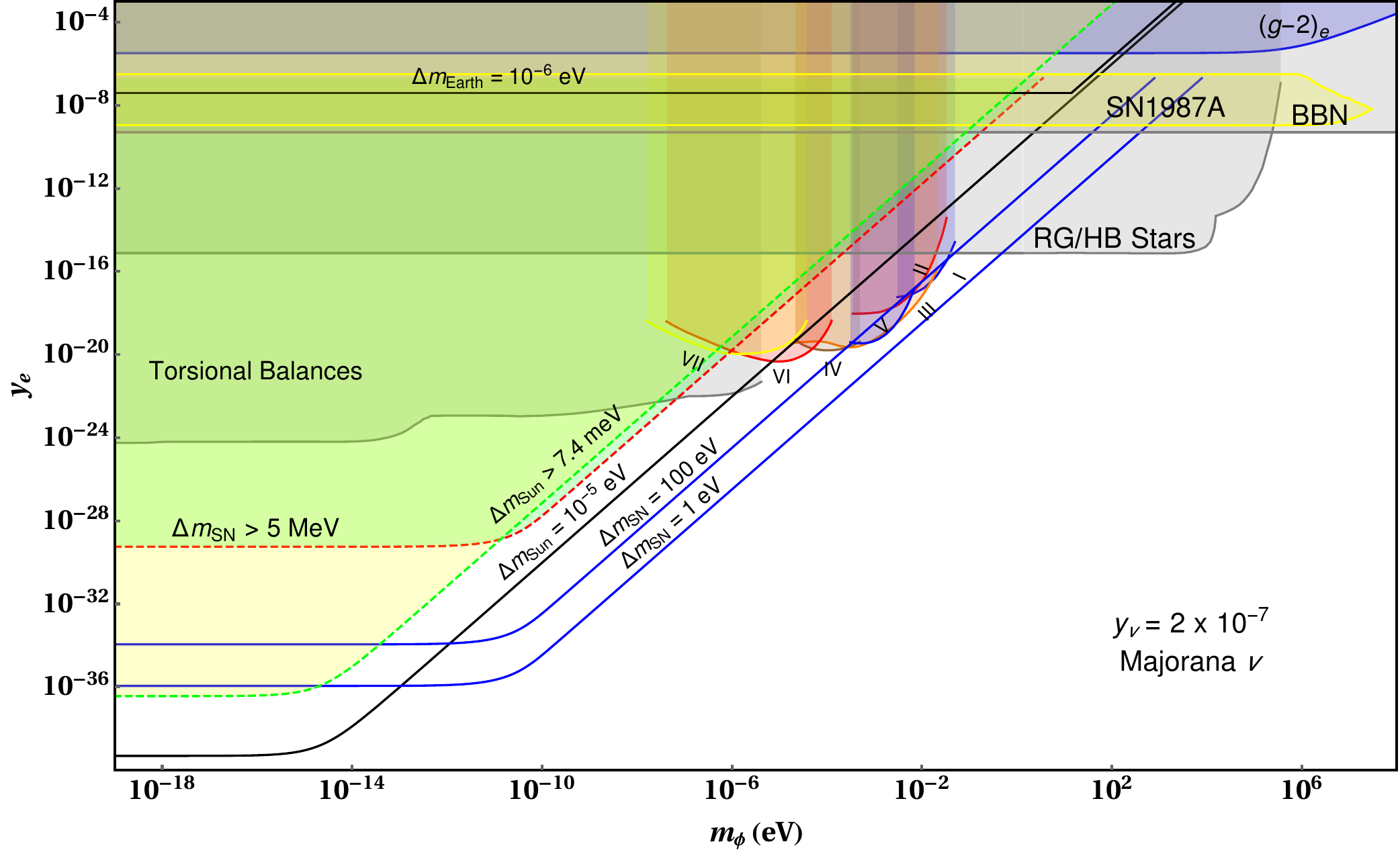}
    \caption{Same as in Fig.~\ref{fig:YeD}, but for Majorana neutrinos.}
    \label{fig:YeM}

  \vspace*{\floatsep}

  \includegraphics[width=0.98\textwidth]{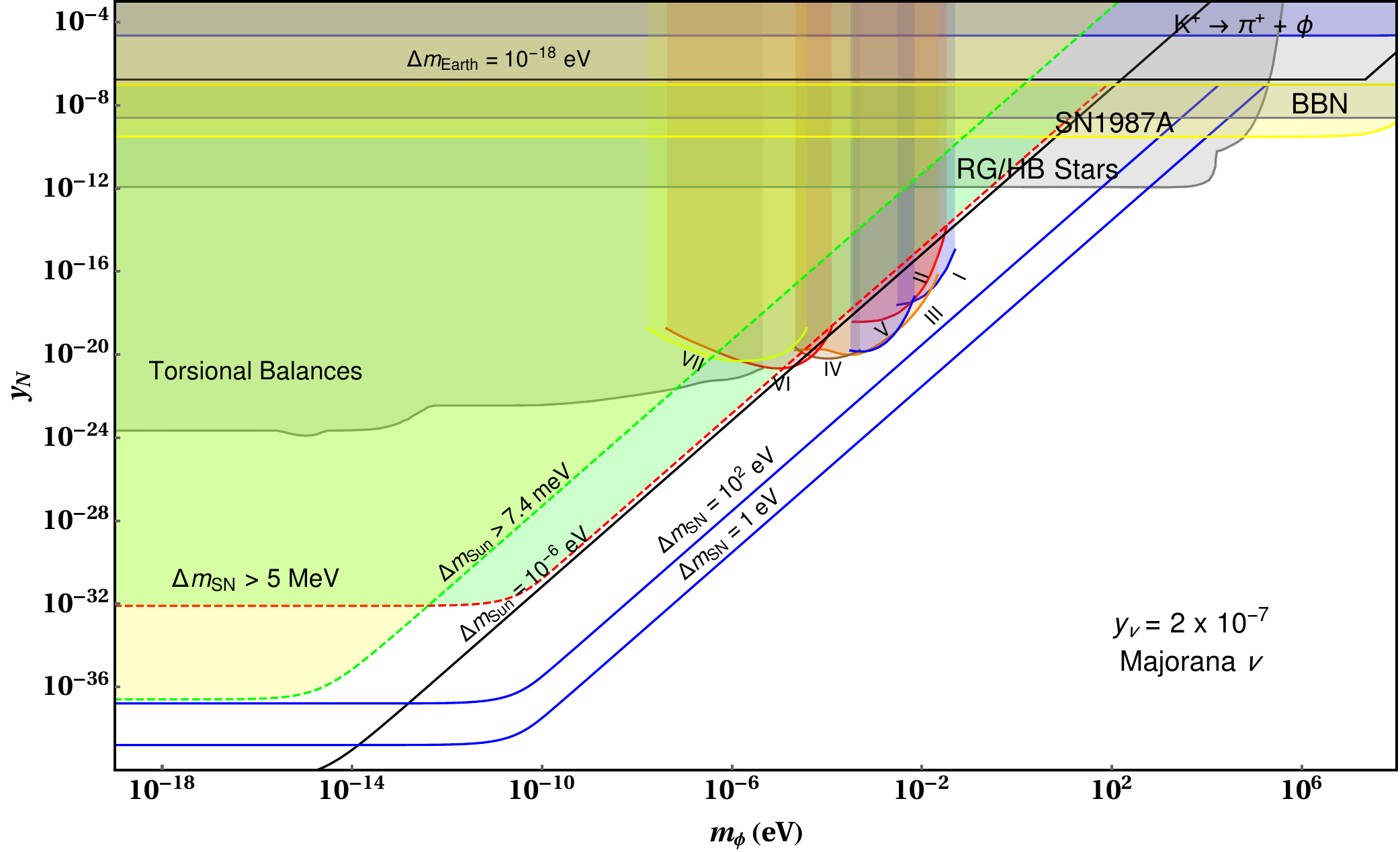}
    \caption{Same as in Fig.~\ref{fig:YnD}, but for Majorana neutrinos.}
    \label{fig:YnM}
\end{figure}

We have discussed the calculation for scalar NSI and the experimental constraints on them in previous sections. Here we put these constraints together and explore possible tests of this scenario in future neutrino experiments.  We also provide the numerical models for the density profiles of the earth and supernovae that we adopt to constrain the model parameters. 

The results for different cases with scalar coupling to electron/nucleon and in case of either Dirac or Majorana neutrinos have been presented in Figs.~\ref{fig:YeD}, \ref{fig:YnD}, \ref{fig:YeM}, \ref{fig:YnM}. Here we have fixed the value of $y_\nu$ at its maximum allowed value in each case, as discussed in Sec.~\ref{sec:4.2}, whereas the other Yukawa coupling (either $y_e$ or $y_N$) is varied, along with the scalar mass $m_\phi$. These results are also summarized in Table \ref{table:NSIValues}.

\subsection{Earth and Sun}
In case of Earth and Sun, the background medium of electrons and nucleons are non-relativistic. 
Therefore, the expression used for scalar NSI in these media is given by Eq.~\eqref{eq:c1} with $N_{\bar f}=0$:
\begin{equation}
    \Delta m_{\nu,{\alpha\beta}} \  = \ \frac{y_f y_{\alpha\beta}}{m_\phi^2}N_f~.
    \label{eq:7.1}
\end{equation}
From the discussion in Sec.~\ref{sec:QM}, when the mediator mass becomes lower than the quantum mechanical cut-off of $m_0\sim 14$ eV, $m_0^2$ should be used in the denominator of Eq.~\eqref{eq:7.1} in lieu of $m_\phi^2$ for Earth. This leads to the turning of the scalar NSI line in the plots for Earth. We have used $N_e^\text{Earth}= 5.4\: N_A \text{ cm}^{-3}$ \cite{Tanabashi:2018oca} and $N_{N}^\text{Earth}= \frac{2.9}{m_N}\:  \text{ g} \text{ cm}^{-3}$ \cite{PREM}, where the nucleon mass $m_N=931.5$ MeV and the Avogadro number $N_A=6.022 \times 10^{23}$. As can be seen from the plots, there are no prospects for observable scalar NSI to be detected on Earth in any of the four cases (Dirac/Majorana and coupling to electrons/nucleons). It can be seen from Table. \ref{table:NSIValues} that highest allowed value of  scalar NSI in case of Earth is around $10^{-14}$ eV for the case of $\phi$ coupling to Majorana neutrinos and electrons. 

\par For the case of Sun, there will also be correction to the scalar NSI from finite size of the medium in the case of light mediators masses $m_\phi \simeq R_{\rm Sun}^{-1}$ as discussed in Sec.~\ref{sec:LongForce} and Ref.~\cite{Smirnov_2019}. We calculate the form factor for Sun using Eq.~\eqref{eq:NSILongR} and the number density of electrons/nucleons, which is obtained by fitting the known solar density profile given in Refs.~\cite{SolarM,Bahcall:1998wm,Bahcall_2005}. We have used the following best fit to the number density profile for Sun:
\begin{align}
     N(r)_{\text{e}} & \ = \ 111.61 N_A \; e^{-(4.81\; r + 10.21\: r^2)} \text{ cm}^{-3} \qquad& \text{(for electron)} \, , \\
    N(r)_{N} & \ = \ \frac{157.13}{m_N}\; e^{-(6.1\; r + 5.2\: r^2)}  \text{ g} \text{ cm}^{-3} & \text{(for nucleon)} \, .
\end{align}

\par As can be seen from the plots, the existing laboratory and astrophysical constraints do allow for a non-negligible scalar NSI in the Sun, especially for $m_\phi\lesssim 1 \mu$eV where the NSI can be as large as  $10^5$ eV for the case of $\phi$ coupling to Dirac/Majorana neutrinos and electrons. However, this will lead to a large correction term to the solar neutrino mass, which is severely constrained by solar neutrino data. Using the $\chi^2$-analysis of the Borexino data from Ref.~\cite{Ge:2018uhz}, we find a $3\sigma$ upper bound on the scalar NSI in Sun: $\Delta m_{\rm Sun}\lesssim 7.4\times 10^{-3}$ eV, as shown by the yellow shaded region in Figs.~\ref{fig:YeD}, \ref{fig:YnD}, \ref{fig:YeM}, and  \ref{fig:YnM}. This still leaves some room for observable scalar NSI effects in future solar neutrino data, especially for ultra-light scalar mediators. Note that very small coupling values for which $y_f^2\lesssim G m^2_\nu = (m_\nu/M_{\rm Pl})^2 \sim 10^{-30}$ are disfavored by the weak gravity conjecture~\cite{ArkaniHamed:2006dz} which suggests gravity as the weakest force in nature.

\subsection{Supernovae}

In the case of supernovae with a typical core temperature $T\sim 30$ MeV, the electron background is relativistic while the nucleon background can be essentially treated to be at rest. Thus, there are two different expressions to be used [cf.~Eqs.~\eqref{eq:c1} and \eqref{eq:c2}]:
\begin{numcases}{
 \Delta m_{\nu,{\alpha\beta}}  \ = \ } 
\frac{y_f y_{\alpha\beta}}{m_\phi^2}N_{N}^{\rm SN}  & \text{(for nucleon)}  \\
  \frac{y_{\alpha\beta}y_f}{m_\phi^2}\frac{m_e}{2}\left(\frac{3N_{e}^{\rm SN}}{\pi}\right)^\frac{2}{3} &  \text{(for electron)}~.
  \end{numcases}
Similar to the case in Sun, there will be correction to the scalar NSI in supernova from the finite size of the medium. Therefore, we numerically integrate Eq.~\eqref{eq:NSILongR} to obtain the form factor for a realistic supernova density profile. We use the fiducial model parameters from Ref.~\cite{Chang_2017} given below: 
\begin{numcases}{\rho(r) \ = \ \rho_c \times}
1+k_\rho(1-r/R_c)\; & \textbf{     ($r < R_c$)}  \\
(r/R_c)^{-\eta}  & \textbf{     ($r \geq R_c$)}
\end{numcases}
where $\rho_c=3 \times 10^{14}  \text{ g} \text{ cm}^{-3}$ is the density at core radius $R_c=10$ km , $k_\rho=0.2$ and $\eta=5$. Assuming the medium to be electrically neutral and using a proton fraction $Y_p=0.3$, we can obtain the number density for electrons from $\rho(r)$.
\par An interesting feature emerges for scalar NSI in a supernova. Due to the high temperature, a light scalar might develop a considerable thermal mass if it has strong enough coupling to the background as discussed in Sec.~\ref{sec:TM}. This leads to Eq.~\eqref{eq:thSNSI} which is independent of $m_\phi$. Trapping leads to the thermalization of the scalar in the medium. Thus, we have only plotted the scalar NSI expression for the supernova as long as it is not trapped inside. 
\par Scalar NSI produced in a supernova cannot be arbitrarily high. If it becomes too large, then neutrino production would be affected in direct conflict with observations from SN1987A. For typical supernova core temperature around $T\simeq 30$ MeV, we constrain the scalar NSI to be less than $5$ MeV \cite{Smirnov_2019}, so that neutrinos around $10$ MeV could be detected on Earth from SN1987A. In the plots, this bound is shown as a dashed line marked $\Delta m_{\rm SN} > 5$ MeV. In any case, we find that sizable scalar NSI can still be observed in supernovae, while being consistent with all other constraints. 

\begin{table}
\centering
\begin{tabular}{ |c|c|c|c| } 
\hline
Case & \small{Max. NSI (eV)}  & \small{Scalar Mass Range (eV)} & Range for $y_f$ \\
\hline
\textbf{Dirac $\nu$, $\phi-e$} & & & \\
Earth & $3.0 \times 10^{-17}$ & 0.04\:-14 & $\sim 7.0 \times 10^{-16}$ \\
Sun & $7.4 \times 10^{-3}$ & $<  10^{-11}$ & $3.3 \times 10^{-34}\:- 10^{-26}$ \\
Supernova & $5.0\times 10^{6}$ & $10^{-11}\:-  10^{-9}$ & $ 10^{-26}\:- 1.8 \times 10^{-23}$ \\
\hline
\textbf{Dirac $\nu$, $\phi-N$} & & & \\
Earth & $10^{-24}$ &$5.3 \times 10^{3}\:- 2.1 \times 10^{7}$ & $\sim 2.4 \times 10^{-10}$ \\
Sun & $7.4 \times 10^{-3}$ & $< 3.3 \times 10^{-13}$ & $2.4\times 10^{-34}\:- 7.5\times 10^{-30}$ \\
Supernova & $5.0\times 10^{6}$ & $3.3\times 10^{-13}\:- 1.8 \times 10^{-7}$ & $7.5 \times 10^{-30}\:- 4.9 \times 10^{-22}$ \\
\hline
\textbf{Majorana $\nu$, $\phi-e$} & & & \\
Earth & $10^{-14}$ & 0.04\:-14 & $\sim 6.0 \times 10^{-16}$ \\
Sun & $7.4 \times 10^{-3}$ &  $<  10^{-11}$ & $4.4\times 10^{-37}\:- 8.7\times 10^{-30}$ \\
Supernova & $5.0\times 10^{6}$ & $10^{-11}\:- 7 \times 10^{-8}$ & $8.7\times 10^{-30}\:- 9.3 \times 10^{-23}$ \\
\hline
\textbf{Majorana $\nu$, $\phi-N$} & & & \\
Earth & $10^{-21}$ &$5.3 \times 10^{3}\:- 2.1 \times 10^{7}$ & $\sim 2.1 \times 10^{-10}$ \\
Sun & $7.4 \times 10^{-3}$ & $< 3.5 \times 10^{-13}$ & $3.1\times 10^{-37}\:- 8.4 \times 10^{-33}$ \\
Supernova & $5.0\times 10^{6}$ & $3.5 \times 10^{-13}\:- 1.3\times 10^{-5}$ & $8.4 \times 10^{-33}\:- 2.0 \times 10^{-21}$ \\
\hline
\end{tabular}
\caption{The maximum allowed value of scalar NSI in different cases and domains with corresponding ranges for the scalar mass $\phi$ and the coupling strength $y_f$, for a fixed $y_\nu$ as shown in Figures~\ref{fig:YeD}-\ref{fig:YnM}.}
\label{table:NSIValues}
\end{table}

\section{UV-complete model for scalar NSI} \label{sec:8}

In this section, we sketch possible ultraviolet completions that would induce interactions of neutrinos with a light scalar.  This discussion is intended only as a proof of principle.  We focus on the case of Dirac neutrinos, with a light scalar $\phi$ coupling to the neutrinos and the electron.

First we construct two effective operators that are invariant under the SM gauge symmetry.  One induces couplings of the scalar $\phi$ to neutrinos and the other to the electron.  These operators are

\begin{equation}
(i)~~~~ \overline{\psi}_L\widetilde{H}\nu_R\frac{\phi}{\Lambda_\nu},~~~~~~~ (ii)~~~~ \overline{\psi}_L H e_R\frac{\phi^2}{\Lambda_e^2}~.
\label{operators}
\end{equation}
Here $\phi$  is a real scalar field, which is a singlet under SM symmetry, $H=\big(\begin{smallmatrix}
  H^+ \\
  H^0
\end{smallmatrix}\big)$ is the SM Higgs doublet and $\psi_L=\big(\begin{smallmatrix}
  \nu \\
  e
\end{smallmatrix}\big)_L$ is the left-handed lepton doublet. These effective operators exhibit a $Z_2$ symmetry (apart from lepton number) under which $\nu_R$ and $\phi$ are odd, with other fields being even.  $\phi$ develops a vacuum expectation value, $\langle \phi \rangle = v_\phi \sim 10$ eV, which breaks the $Z_2$ symmetry.  The neutrino Yukawa coupling $y_\nu$ and the electron Yukawa coupling $y_e$ with the $\phi$ field are respectively given by
\begin{equation}
y_\nu \ = \ \frac{v}{\Lambda_\nu}\, , \qquad y_e \ = \  \frac{2 v v_\phi}{\Lambda_e^2}
\end{equation}
where $v= 174$ GeV is the VEV of the SM Higgs doublet.  Once $\phi$ acquires a VEV, the operator $(i)$ generates a mass term for the neutrino given by
\begin{equation}
    m_\nu \ = \ \frac{v_\phi v}{\Lambda_\nu}~.
\end{equation}
While this may be the leading contribution, it is not required to be so, as there could be other contributions as well.  In any case, this would imply an upper limit on $y_\nu$ given by
\begin{equation}
    y_\nu \ < \ \frac{m_\nu}{v_\phi}~.
    \label{mnu}
\end{equation}
The cut-off scale $\Lambda_e$ is expected to be at least a  hundred GeV, while $\Lambda_\nu$ may be lower.  Choosing $\Lambda_e \sim v$, we would have $y_e \sim v_\phi/v$.  For $y_e \sim 10^{-10}$, as our analysis requires for observable scalar NSI, $v_\phi \sim 10$ eV is preferable.  This in turn implies from Eq.~(\ref{mnu}) that $y_\nu < 5 \times 10^{-3}$, using $m_\nu \equiv \sqrt{\Delta m^2_{\rm atm}} \sim 0.05$ eV. $y_\nu$ of course can be smaller than this value, which would be in the  interesting range for observable scalar NSI.  

\begin{figure}[t!]
    \centering
    \includegraphics[scale=1.3]{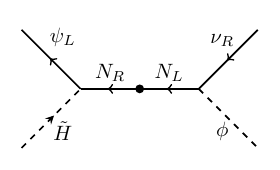} 
      \includegraphics[scale=1.3]{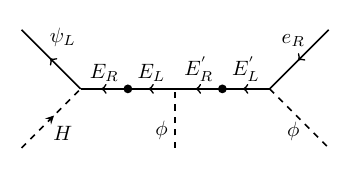}
    \caption{Explicit models generating operators of Eq.~(\ref{operators}).}
    \label{fig:caseA}
\end{figure}

The operators in Eq.~(\ref{operators}) can be generated by adding new vector-like fermions to the SM.  For example, operator $(i)$ can arise by the addition of SM singlet fermions $N_{L,R}$ with a lepton number preserving Dirac mass.  The relevant Lagrangian is given by
\begin{equation}
    \mathcal{L} \ \supset \ y_N\overline{\psi}_L\widetilde{H} N_R + M_N \overline{N}_R N_L + y^\nu_\phi \overline{N}_L \nu_R \phi + {\rm H.c.}
\end{equation}
These interactions also preserve the $Z_2$ symmetry with $N_{L,R}$ being even under it. The diagram generating operator $(i)$ is shown in Fig.~\ref{fig:caseA}, left panel. 

Operator $(ii)$ is induced by integrating out a pair of vector-like leptons, $E,\,E'$, both being singlets of $SU(2)_L$ and carrying hypercharge $Y = -2$. Their interaction Lagrangian is given by
\begin{equation}
    \mathcal{L} \ \supset \ y_E\overline{\psi}_L {H} E_R + \mu_E \overline{E}_R E_L  + y^E_\phi \overline{E}_L E^{'}_R \phi +  M_{E^{'}} \overline{E}_R^{'} E_L + y^e_\phi \overline{E}_L^{'} e_R \phi + {\rm H.c.}
\end{equation}
Here $E_{L,R}$ are even and $E_{L,R}^{'}$ are odd under $Z_2$. The effective operator involving electron and $\phi$ is generated by Fig.~\ref{fig:caseA}, right panel.

Integrating out the heavy degrees of freedom we obtain the following effective Lagrangian terms:
\begin{equation}
(i) ~~~\frac{y_N y^\nu_\phi}{M_N} \overline{\psi}_L\widetilde{H} \nu_R \phi,~~~~~~~
(ii)~~~ \frac{y_E y^E_\phi y^e_\phi}{M_E M_{E^{'}}} \overline{\psi}_L {H} e_R \phi^2~.
\end{equation}
These expressions can be mapped to Eq.~(\ref{operators}) to identify the cut-off scales $\Lambda_\nu$ and $\Lambda_e$, and the constraints discussed in terms of the cut-off scales will apply to them. We thus see broad consistency of the model.  In particular, the induced neutrino mass from these interactions is not excessive and the vector-like leptons having mass of order few hundred GeV is consistent with collider data.  Note that breaking the $Z_2$ at a scale of order 10 eV does not cause cosmological domain wall problem, since the energy density carried by the walls is quite small.  We have ignored here possible mixing between the $\phi$ and $H$ fields since such mixing is small, of order $v_\phi/v$ and is  controlled by a new quartic coupling which may also be small.

\section{Conclusion} \label{sec:9}
We have performed a systematic study of scalar NSI of neutrinos with matter due to a light scalar mediator. First, a general field-theoretic derivation of the scalar NSI formula is given, which is valid at arbitrary temperature and density, and hence, applicable in widely different environments, such as Earth, Sun, supernovae and early Universe. We have also extended the analysis of long-range force effects for all background media, including both relativistic and non-relativistic limits. Using these results and applying various experimental and astrophysical constraints, we find that observable scalar NSI has been precluded in terrestrial experiments, primarily due to atomic form factor effects, which can also be understood from simple quantum-mechanical uncertainty principle. Nevertheless, sizable scalar NSI effects are still possible in the Sun, supernovae and early Universe environments, which could be detected in future solar and supernova neutrino data, as well as in the form of extra relativistic species ($\Delta N_{\rm eff}$) and neutrino self-interactions in cosmological observations. We have also presented examples of UV-complete models that could give rise to such scalar NSI effects.

\section*{Acknowledgments}

We thank Mark Alford, Steven Harris, Ahmed Ismail   and Pedro Machado for discussions. This work was supported in part by US Department of Energy
Grant Numbers DE-SC 0016013 (K.S.B.) and DE-SC0017987 (B.D., G. C.). This work was also supported by the Neutrino Theory Network Program under Grant No. DE-AC02-07CH11359. K.S.B. and B.D. thank the Fermilab Theory Group for warm hospitality, where part of this work was done. B.D. and G.C. also thank the Department of Physics at Oklahoma State University for warm hospitality during the completion of this work.

\appendix
\section*{Appendices}

\section{Limiting cases for scalar NSI expression}
\label{appendix:1}
In this Appendix we evaluate the self-energy given in Eq.~\eqref{eq:sNSI} corresponding to the tadpole diagram of Fig.~\ref{fig:FD}.  We shall evaluate only the fermionic contribution to Eq.~\eqref{eq:sNSI}, from which it is easy to read of the anti-fermionic background contribution as well. We also provide an exact expression for the medium-dependent neutrino mass, which can be evaluated numerically.

\subsection{Case 1: $\mu > m_f \gg T$}
Breaking the integration limits and expanding the occupation number as an infinite series, we can write Eq.~\eqref{eq:sNSI} as follows:
\begin{align}
     \Delta m_{\nu, {\alpha\beta}} & \ = \ \frac{ m_fy_{\alpha\beta}y_f}{2 \pi^2 m_\phi^2}\left(\left[\mu\sqrt{\mu^2-m_f^2}+m_f^2\ln{\left(\frac{m_f}{\mu+\sqrt{\mu^2-m_f^2}}\right)} \right] \right. \nonumber\\
     & + \left. \sum_{n=1}^\infty (-1)^n \left[ \int_{m_f}^\mu dE\:e^{n(E-\mu)/T}\sqrt{E^2-m_f^2} + \int_{\mu}^\infty dE\:e^{-n(E-\mu)/T}\sqrt{E^2-m_f^2}\right] \right)~.
\end{align}
As $T\rightarrow0$, the first term in the series dominates the result. We know that sum over all momentum states weighted by occupation number yields the number density. Inverting the relation to obtain $\mu$, we get:
\begin{equation}
    \mu^2 \ = 
    \ (3\pi^2 N_f)^\frac{2}{3} + m_f^2 \ \simeq \  (3\pi^2 N_f)^\frac{2}{3},
\end{equation}
where in the second relation we assumed $\mu^2 \gg m_f^2$.
Thus, for $\mu \gg m_f$ we have
\begin{equation}
    \Delta m_{\nu, {\alpha\beta}} \  \simeq \  \frac{y_{\alpha\beta}y_f}{m_\phi^2}\frac{m_f}{2}\left(\frac{3N_f}{\pi}\right)^\frac{2}{3}~,
\end{equation}
as given in Eq.~\eqref{eq:c2}. 

\subsection{Case 2: $T \ll \mu < m_f$}
When $\mu<m_f$, the expression for $\Sigma$ of Eq.~(\ref{eq:sNSI}) can be written as a weighted series of modified Bessel function of the second kind:
\begin{equation}\label{eq:temp1}
    \Delta m_{\nu,{\alpha\beta}} \ = \ \frac{ m_fy_{\alpha\beta}y_f}{ \pi^2 m_\phi^2} \sum_{n=1}^\infty (-1)^{n+1}\:\frac{m_fT}{n} e^{n\mu /T} K_1\left(\frac{nm_f}{T}\right) \, .
\end{equation}
For $z\rightarrow \infty$, we can use the asymptotic form for $ K_\nu\left(z\right)$ :
\begin{equation}\label{eq:KLowT}
    K_\nu\left(z\right) \ \simeq \ e^{-z} \sqrt{\frac{\pi }{2z}}\left(1+ \frac{4 \nu^2-1}{8z} + ... \right)~.
\end{equation}
Due to the exponential suppression, the $n=1$ term in the sum will be dominant in Eq.~\eqref{eq:temp1}. This yields:
\begin{equation}\label{eq:XlmlT}
    \Delta m_{\nu,{\alpha\beta}} \ \simeq \ \frac{2 y_f y_{\alpha\beta}}{m_\phi^2}\left( \frac{m_f T}{2 \pi}\right)^{\frac{3}{2}}e^{-(m_f-\mu)/T}~.
\end{equation}
To relate the above function to the number density $N_f$, we use
\begin{align}\label{eq:Nd}
    N_f & \ = \ 2 \int \frac{d^3k}{(2\pi)^3} \frac{1}{e^{(E-\mu)/T}+1} \nonumber  \\
    & \ = \ \frac{1}{\pi^2}\int_{m_f}^\infty dE\:\frac{E\sqrt{E^2-m_f^2}}{e^{(E-\mu)/T}+1} \nonumber  \\
    & \ = \  \frac{1}{\pi^2}\sum_{n=1}^\infty\int_{m_f}^\infty dE\:E\sqrt{E^2-m_f^2}\:e^{-n(E-\mu)/T} (-1)^{n+1} \nonumber \\
    & \ = \ \frac{1}{ \pi^2} \sum_{n=1}^\infty (-1)^{n+1}\:\frac{m_f^2T}{n} e^{n\mu /T} K_2\left(\frac{nm_f}{T}\right)~.
\end{align}
Using Eq.~\eqref{eq:KLowT} in the expression above and retaining only the dominant $n=1$ term, we have 
\begin{equation}
    N_f \ \simeq \ 2 \left( \frac{m_f T}{2 \pi}\right)^{\frac{3}{2}}e^{-(m_f-\mu)/T} ~.
\end{equation}
Thus, the medium-induced neutrino mass in the limit $T \ll \mu<m_f$ evaluates to:
\begin{equation}\label{eq:PG}
    \Delta m_{\nu,{\alpha\beta}} \ \simeq \  \frac{y_f y_{\alpha\beta}}{m_\phi^2}N_f\, , 
\end{equation}
as given in Eq.~\eqref{eq:c1}. 

\subsection{Case 3: $\mu < m_f \ll T$}

For $z\rightarrow 0$, the asymptotic form for $ K_\nu\left(z\right)$ is:
\begin{equation}\label{eq:KhighT}
    K_\nu\left(z\right) \ \simeq \  \frac{\Gamma(\nu)}{2} \left(\frac{z }{2}\right)^{-\nu}~.
\end{equation}
Using the above in Eq.~\eqref{eq:temp1}, we can write the mass correction as:
\begin{align}
     \Delta m_{\nu,{\alpha\beta}} & \ \simeq \  \frac{ m_fy_{\alpha\beta}y_f}{ \pi^2 m_\phi^2} \sum_{n=1}^\infty (-1)^{n+1}\:\frac{T^2}{n^2} e^{n\mu /T} \\
     & \ = \ -\frac{ m_fy_{\alpha\beta}y_fT^2}{ \pi^2 m_\phi^2}\text{Li}_2(-e^{\mu/T}) \, ,
\end{align}
where $\text{Li}_\nu(z)$ is the polylogarithm. In the case $|z|\rightarrow0$,  $\text{Li}_n(-e^{z }) \simeq -(1-2^{1-n})\zeta(n)$. Using this one   obtains:
\begin{equation}
   \Delta m_{\nu,{\alpha\beta}} \ \simeq \ \frac{y_f y_{\alpha\beta}m_f T^2}{12 m_\phi^2} ~.
\end{equation}
Again using Eq.~\eqref{eq:KhighT} in Eq.~\eqref{eq:Nd} and retaining only the $n=1$ term we get: 
\begin{align}
    N_f \ \simeq \ -\frac{2T^3}{\pi^2} \text{Li}_3(-e^{\mu/T}) \ = \ \frac{3T^3}{2\pi^2}\zeta(3)~.
    \label{eq:A14}
\end{align}
Thus, the scalar NSI expression for $\mu<m_f \ll T$ evaluates to:
\begin{equation}\label{eq:PG}
    \Delta m_{\nu,{\alpha\beta}} \ \simeq \  \frac{y_{\alpha\beta}y_fm_f}{3 \:m_\phi^2}\left(\frac{\pi^2N_f}{12 \:\zeta(3)}\right)^\frac{2}{3}~,
\end{equation}
as given in Eq.~\eqref{eq:c3}.

\section{Calculation of neutrino self-energy in neutrino background}\label{appendix:3}
Here we evaluate the neutrino self-energy arising from a neutrino background as given in Eq.~\eqref{eq:calcSE}. 
We can rewrite the delta function in Eq.~\eqref{eq:calcSE} as follows:
\begin{align}
    \delta\left[{\left(k+\frac{p}{2}\right)^2-m_\phi^2}\right] & \ = \  \frac{1}{|\bold{k}||\bold{p}|}\delta(\cos{\theta}-\cos{\theta_0}) \, ,
\end{align}
where
\begin{equation}
    \cos{\theta_0} \ = \ \frac{k_0^2-|\bold{k}|^2+\frac{p^2}{4}-m_\phi^2+k_0p_0}{|\bold{k}||\bold{p}|}~.
\end{equation}
Using kinematical arguments and $|\cos{\theta_0}|\leq 1$, we find the range for $k_0$ and $|\bold{k}|^2$:
\begin{equation}
    k_0 \ : \ \left\{\frac{-p_0}{2}+m_\nu,\infty \right\},~~~~
    |\bold{k}|^2  \ : \ \left\{|\bold{k}|^2_-\;, \:|\bold{k}|^2_+ \right\}
\end{equation}
where
\begin{equation}
    |\bold{k}|^2_\pm \ = \ \frac{1}{4}\bigg(|(\bold{p}|\pm\sqrt{|(\bold{p}|)^2+4k_0p_0-4m_\nu^2+4k_0^2+p^2}\bigg)^2~.
\end{equation}
Changing the integration variables to spherical coordinates and integrating over  $\cos{\theta}$ we obtain:
\begin{equation}
    \Sigma_{\alpha\beta}^\nu \ = \  -\frac{y_{\alpha\gamma} y_{\gamma\beta}}{16\pi^2|\textbf{p}|} \int_{k_0^\textbf{min}}^{k_0^\textbf{max}} dk_0 \int_{|\bold{k}|^2_-}^{|\bold{k}|^2_+} d|\textbf{k}|^2\:\frac{(\slashed{k}+\frac{\slashed{p}}{2}+m_\nu)}{k_0^2-|\textbf{k}|^2+\frac{p^2}{4}-\frac{m_\phi^2+m_\nu^2}{2}}\:n_\nu \left(k_0+\frac{p_0}{2}\right)~.
\end{equation}
This contribution can be decomposed as given in Eq.~\eqref{eq:2.19}. 
By defining
\begin{equation}
    \mathcal{I} \ = \ \int_{m_\nu}^\infty dk_0\; n_\nu(k_0)\ln\left[\frac{k_0p_0-p^2+\frac{m_\phi^2-m_\nu^2}{2}+|\textbf{p}|\sqrt{k_0^2-m_\nu^2}}{k_0p_0-p^2+\frac{m_\phi^2-m_\nu^2}{2}-|\textbf{p}|\sqrt{k_0^2-m_\nu^2}}\right]\, ,
    \label{eq:B6}
\end{equation}
the quantities $J_u,J_m,J_p$ in Eq.~\eqref{eq:Jpum} can be written succinctly as:
\begin{align}
    J_m \ & = \ -2 m_\nu \:\mathcal{I} \, ,\\
    J_p \ & = \ -(p^2+m_\nu^2-m_\phi^2)\:\mathcal{I} -2 |\textbf{p}|\int_{m_\nu}^\infty dk_0\; n_\nu(k_0)\sqrt{k_0^2-m_\nu^2} \, ,\\
    J_u \ & = \ -2\int_{m_\nu}^\infty dk_0\;k_0\; n_\nu(k_0)\ln\left[\frac{k_0p_0-p^2+\frac{m_\phi^2-m_\nu^2}{2}+|\textbf{p}|\sqrt{k_0^2-m_\nu^2}}{k_0p_0-p^2+\frac{m_\phi^2-m_\nu^2}{2}-|\textbf{p}|\sqrt{k_0^2-m_\nu^2}}\right]~. \label{eq:B9}
\end{align}
These integrals $(J_m,\,J_p,\, J_u)$ cannot be evaluated analytically in general.  However, they may be evaluated in the high temperature limit.  For this purpose we set $m_\nu$ to zero and assume the chemical potential $\mu$ is small.  This condition should be realized when the results are applied to early Universe. The integrals in this limit are evaluated to be:
\begin{eqnarray}
J_m & \ \simeq \ & -2 m_\nu T\, {\rm ln} 2 \,\,{\rm ln}\left(\frac{2 \sqrt{2} |{\bf p}|T}{m_\phi^2} \right) \, ,  \\
J_p & \ \simeq 
\ & \frac{\pi^2 T^2|{\bf p}|}{3} + |{\bf p}|^2 T\, {\rm ln}2 \,\, {\rm ln}\left(\frac{2 \sqrt{2} |{\bf p}| T}{m_\phi^2} \right) \, , \\
J_u & \ \simeq \ & \frac{\pi^2 T^2}{6}\left(12 \zeta'(-1) + {\rm ln}\left(\frac{16 \pi |{\bf p}| T}{m_\phi^2} \right)  \right)~.
\end{eqnarray}
These results have been applied to derive the energy shift for neutrinos and antineutrinos in Sec.~\ref{sec:2.2}, see Eq.~(\ref{energy-shift}).

Similar calculation can be performed for the case of thermalized scalar field $\phi$. By defining :
\begin{equation}
    \mathcal{I}^\phi \ = \ \int_{m_\phi}^\infty dk_0\; n_\phi(k_0)\ln\left[\frac{k_0p_0+p^2+|\textbf{p}|\sqrt{k_0^2-m_\phi^2}}{k_0p_0+p^2-|\textbf{p}|\sqrt{k_0^2-m_\phi^2}}\right]\, ,
\end{equation}
the contribution from thermal $\phi$ to Eq.~\eqref{eq:Jpum} can be  labeled as $J_m^\phi,J_p^\phi,J_u^\phi$ and given by:
\begin{align}
    J_m^\phi \ & = \ -2 m_\nu \:\mathcal{I}^\phi \, ,\\
    J_p^\phi \ & = \ -p^2\:\mathcal{I}^\phi +2 |\textbf{p}|\int_{m_\phi}^\infty dk_0\; n_\phi(k_0)\sqrt{k_0^2-m_\phi^2} \, ,\\
    J_u^\phi \ & = \ -2\int_{m_\phi}^\infty dk_0\;(k_0+p_0)\; n_\phi(k_0)\ln\left[\frac{k_0p_0+p^2+|\textbf{p}|\sqrt{k_0^2-m_\phi^2}}{k_0p_0+p^2-|\textbf{p}|\sqrt{k_0^2-m_\phi^2}}\right]~. 
\end{align}
These terms should be added to the terms $J_p,\,J_u,\,J_m$ of Eq.~(\ref{eq:Jpum}) so that they become $J_p + J_p^\phi,\,  J_u + J_u^\phi,\, J_m + J_m^\phi$.  The results of the matter-dependent neutrino mass will go through with these replacements. 

\section{Examples for finite medium effects in relativistic cases}
\label{appendix:4}

Here we work out Eq.~\eqref{eq:NSILongR} in the relativistic limit for two different density profile distributions.
\subsection{Constant density distribution}
For a relativistic medium like electron background in supernovae, the quantity $\langle \bar{f} f \rangle $ in Eq.~\eqref{eq:NSILongR} takes the form:
\begin{equation}
    \langle \bar{f} f \rangle_{\text{SN}} \ = \   \frac{m_f}{2}\left(\frac{3 N_f}{\pi}\right)^\frac{2}{3} \, .
\end{equation}
Consider a constant density distribution such that 
\begin{equation}
    N_f(r) \ = \  N_f(0)\:\Theta(R-r) \, ,
\end{equation}
where $R$ is the radius of the constant-density spherical body. Plugging the $\langle \bar{f} f \rangle $ in Eq.~\eqref{eq:NSILongR} yields a general form for scalar NSI in relativistic media with $\mu>m_f\gg T$ :
\begin{align}\label{eq:relFF}
    \Delta m_{\nu,{\alpha\beta}}(r) \ = \   \frac{y_{\alpha\beta}\:y_f}{m_\phi \: r} \frac{m_f}{2}  \left(\frac{3}{\pi}\right)^\frac{2}{3} & \left( e^{-m_\phi r}\int_0^r x  \: N_f^{2/3} \: \sinh{(m_\phi\: x)}\: dx \right. \nonumber \\ & \left. + \sinh{(m_\phi \: r)} \int_r^\infty x  \: N_f^{2/3} \: e^{-m_\phi\: x} \: dx  \right)~.
\end{align}
For number density profile in consideration, the above equation yields:
\begin{numcases}{\Delta m_{\nu, {\alpha\beta}}(r) \ = \  \frac{y_{\alpha\beta}\:y_fm_f}{2m_\phi \: r}\left(\frac{3N_f(0)}{\pi}\right)^\frac{2}{3} \times }
 F_< & \textbf{     ($r \leq R$)} \, ,  \\
 F_>& \textbf{     ($r > R$)} \, ,
\end{numcases}
where
\begin{align}\label{eq:Fin}
    F_< \ & = \ 1-\frac{m_\phi R+1}{m_\phi\:r}e^{-m_\phi\:R}\sinh{(m_\phi\:r)} \, , \\
\label{eq:Fout}
    F_> \ & = \ \frac{e^{-m_\phi\:r}}{m_\phi\:r}[m_\phi\:R \cosh{(m_\phi\:R)}-\sinh{(m_\phi\:R)}]~.
\end{align}
Note that the pre-factor in Eq.~\eqref{eq:relFF} matches the scalar NSI contribution calculated in Eq.~\eqref{eq:c2} assuming point contact interaction. 

\par For the non-relativistic case our formalism gives the same result derived in Ref.~\cite{Smirnov_2019} and given below:
\begin{numcases}{\Delta m_{\nu,{\alpha\beta}}(r) \ = \  \frac{y_{\alpha\beta}\:y_f\: N_f(0)}{m_\phi^2} \times }
 F_< & \textbf{     ($r \leq R$)}  \\
 F_>& \textbf{     ($r > R$)}
\end{numcases}
where the functions ($F_<,\:F_>$) are identical to the ones in Eqs.~\eqref{eq:Fin} and \eqref{eq:Fout}.

\subsection{Exponential density distribution}
Given a relativistic medium ($\mu>m_f\gg T$) with the following number density profile:
\begin{equation}
    N_f(r) \ = \ N_f(0)\:e^{-\lambda\:r}\:\Theta(R-r)
\end{equation}
where $R$ is the radius of the spherical  body in consideration, Eq.~\eqref{eq:relFF} yields:
\begin{numcases}{\Delta m_{\nu,{\alpha\beta}}(r) \ = \  \frac{y_{\alpha\beta}\:y_f}{2m_\phi \: r}\left(\frac{3N_f(0)}{\pi}\right)^\frac{2}{3} \times }
 G_< & \textbf{     ($r \leq R$)} \, , \\
 G_>& \textbf{     ($r > R$)} \, ,
\end{numcases}
where
\begin{align}
    G_< \ & = \  \frac{2\lambda m_{\phi}}{3} \left(\frac{e^{m_{\phi}r}\left(\frac{3m_{\phi}^{2}r}{2\lambda}-\frac{2\lambda\:r}{3} -2\right)+2e^{\frac{2\lambda\:r}{3}}}{(m_{\phi}^{2}-\frac{4\lambda^2}{9})^{2}}\right) e^{-r(\frac{2\lambda}{3}+m_{\phi})} \nonumber \\
 & \qquad - \left(\frac{\sinh(m_{\phi}r)(m_{\phi}R+\frac{2\lambda\:R}{3}+1)}{(m_{\phi}+\frac{2\lambda}{3})^{2}}\right) e^{-R(\frac{2\lambda}{3}+m_{\phi})} \, , \\
 G_> \ & = \  \sinh(m_{\phi}R) \left(\frac{m_{\phi}^{2}(\frac{2\lambda\:R}{3}-1)-\frac{4\lambda^2}{9}(\frac{2\lambda\:R}{3}+1)}{\left(m_{\phi}^{2}-\frac{4\lambda^2}{9}\right)^{2}}\right) e^{-(m_{\phi}r+\frac{2\lambda\:R}{3})} 
 +\frac{4\lambda m_{\phi}}{3\left(m_{\phi}^{2}-\frac{4\lambda^2}{9}\right)^{2}}e^{-m_{\phi}r} \nonumber \\
& \qquad +  \cosh(m_{\phi}R) \left(\frac{m_{\phi}^{3}R-\frac{4\lambda^2\:Rm_{\phi}}{9}-\frac{4\lambda\:m_{\phi}}{3} }{\left(m_{\phi}^{2}-\frac{4\lambda^2}{9}\right)^{2}}\right) e^{-(m_{\phi}r+\frac{2\lambda\:R}{3})}~. 
\end{align}
Similar analyses can be done for other relativistic cases such as for early Universe cosmology ($\mu<m_f<T$) albeit with a different pre-factor.
\par For an exponential density distribution with a cut-off in the non-relativistic case we obtain:
\begin{numcases}{\Delta m_{\nu,{\alpha\beta}}(r) \ = \  \frac{y_{\alpha\beta}\:y_f\:N_f(0)}{m_\phi \: r} \times }
 K_< & \textbf{     ($r \leq R$)} \, , \\
 K_>& \textbf{     ($r > R$)} \, ,
\end{numcases}
where we can obtain the functions $K_<$ and $K_>$ by replacing $\lambda\rightarrow\frac{3\lambda}{2}$ in $G_<$ and $G_>$ respectively, i.e., $K(\lambda)_{>(<)}=G(3\lambda/2)_{>(<)}$. This expression is in full agreement with the result of Ref.~\cite{Smirnov_2019}.

\section{Calculation of thermal mass for the scalar field}
\label{appendix:2}

Here we carry out the evaluation of the self-energy diagram of $\phi$ to calculate its thermal mass.  As shown in Sec.~\ref{sec:TM}, $\phi$ can develop a medium-dependent mass, which is given by Eq.~(\ref{eq:ThermalMassPhi}). This contribution can be written as:
\begin{equation}
    \mathcal{M} \ = \ \mathcal{M}_1 + \mathcal{M}_2 \, ,
\end{equation}
where
\begin{align}
    \mathcal{M}_1 \ & = \ 4 y_f^2 \int \frac{d^4p}{(2\pi)^4} \left(k^2-\frac{p^2}{4}+m_f^2\right) \frac{\Gamma_f(k+p/2)}{(k-p/2)^2-m_f^2} \, , \\
    \mathcal{M}_2 \ & = \ 4 y_f^2 \int \frac{d^4p}{(2\pi)^4} \left(k^2-\frac{p^2}{4}+m_f^2\right) \frac{\Gamma_f(k-p/2)}{(k+p/2)^2-m_f^2}~.
\end{align}
Since $\mathcal{M}_1\rightarrow\mathcal{M}_2$ with the replacement $p\rightarrow-p$, we will focus only on simplifying the expression for   $\mathcal{M}_1$.
\begin{equation}
    \mathcal{M}_1 \ = \ 4 y_f^2 \int_{\frac{-p_0}{2}}^\infty dk_0 \int \frac{d^3p}{(2\pi)^3} \left(k^2-\frac{p^2}{4}+m_f^2\right) \frac{\delta((k+p/2)^2-m_f^2)}{(k-p/2)^2-m_f^2}n_f\left(k_0+\frac{p_0}{2}\right) \, .
\end{equation}
The delta function can be written as
\begin{align}
    \delta\left[{\left(k+\frac{p}{2}\right)^2-m_f^2}\right] & \ = \   \frac{1}{|\bold{k}||\bold{p}|}\delta(\cos{\theta}-\cos{\theta_0}) \, ,
\end{align}
where
\begin{equation}
    \cos{\theta_0} \ = \ \frac{k_0^2-|\bold{k}|^2+\frac{p^2}{4}-m_f^2+k_0p_0}{|\bold{k}||\bold{p}|}~.
\end{equation}
Using kinematical arguments and $|\cos{\theta_0}|\leq1$, we find the range for $k_0$ and $|\bold{k}|^2$:
\begin{equation}
    k_0  \ : \ \left\{\frac{-p_0}{2}+m_f,\infty \right\},~~~
    |\bold{k}|^2 \ : \ \left\{|\bold{k}|^2_-\;, \:|\bold{k}|^2_+ \right\} \, ,
    \end{equation}
 where
 \begin{equation}
    |\bold{k}|^2_\pm \ = \ \frac{1}{4}\bigg(|(\bold{p}|\pm\sqrt{|(\bold{p}|)^2+4k_0p_0-4m_f^2+4k_0^2+p^2}\bigg)^2~.
\end{equation}
Thus, changing the integration variables to spherical coordinates and integrating over $\cos{\theta}$ we get:
\begin{equation}
    \mathcal{M}_1 \ = \  -\frac{y_f^2}{4\pi^2|\textbf{p}|} \int_{k_0^\textbf{min}}^{k_0^\textbf{max}} dk_0 \int_{|\bold{k}|^2_-}^{|\bold{k}|^2_+} d|\textbf{k}|^2\:\frac{k_0^2-|\textbf{k}|^2-\frac{p^2}{4}+m_f^2}{k_0^2-|\textbf{k}|^2+\frac{p^2}{4}+m_f^2}\:n_f \left(k_0+\frac{p_0}{2}\right)~.
\end{equation}
Integrating the above integral with respect to  $|\textbf{k}|^2$ and adding the contribution from both $ \mathcal{M}_1$ and $ \mathcal{M}_2$ yields:
\begin{align}
    \mathcal{M} & \ = \   \frac{y_f^2}{\pi^2} \int_{m_f}^\infty dk_0 \:n_f \left(k_0\right)\sqrt{k_0^2-m_f^2}  \nonumber \\
    &  -\frac{y_f^2}{2\pi^2|\textbf{p}|}\left
    ( m_f^2-\frac{m_\phi^2}{4} \right) \int_{m_f}^\infty dk_0\: n_f \left(k_0\right)\:\ln\left(\frac{\left(|\textbf{p}|\sqrt{k_0^2-m_f^2}-\frac{m_\phi^2}{2}\right)^2-k_0^2 p_0^2}{\left(|\textbf{p}|\sqrt{k_0^2-m_f^2}+\frac{m_\phi^2}{2}\right)^2-k_0^2 p_0^2}\right)~.
\end{align}
In the limit $m_\phi \rightarrow 0$, the mass correction for scalar reduces to Eq.~\eqref{eq:ThermalPhi}. 

\bibliographystyle{JHEP}
\bibliography{ScalarNSI}

\end{document}